# PROACT: Parallel Multi-Miner Proof of Accumulated Trust Protocol for Internet of Drones

**Khaleel Mershad** [a,b,*]

[a] Department of Computer Science and Mathematics, Lebanese American University (LAU), Beirut, Lebanon
[b] Department of Computer Science, Modern University for Business and Science (MUBS), Beirut, Lebanon
E-mail: [a] khaleel.mershad@lau.edu.lb, [b] kmershad@mubs.edu.lb

*Abstract*—Several types of networks that comprise unmanned aerial vehicles (UAV or drone) are being utilized in important applications such as emergency response, environment and infrastructure monitoring, defense and security, and commerce. In such networks, swarms of UAVs cooperate in executing one or more missions to achieve the application's objectives. The UAVs communicate with terrestrial networks by connecting to fixed or mobile ground control stations (GCS). The ability of drones to connect to online applications and offer services to Internet users has led to the proliferation of the Internet of Drones (IoD). However, IoD applications are highly vulnerable to many types of cyberattacks. Hence, mechanisms must be deployed to secure the IoD operations and data. Recently, the blockchain has been proposed as a solution to detect and prevent malicious attacks on the UAV network (UAVN). Due to the UAV's limited resources, it becomes a challenge to integrate the blockchain into the IoD. In this paper, we propose a model that enables a drone to store the important data that it requires during its flight within a lightweight blockchain system. In addition, we propose a new blockchain consensus mechanism in which several miners produce their blocks in parallel, which decreases the time needed to add transactions securely to the blockchain and meets the requirements of delay-sensitive applications. Our simulations prove the advantages of the proposed model in decreasing the transaction-to-blockchain delay, the average drone energy consumption, and the blockchain block size as compared to other IoD blockchain systems.

* Corresponding Author

1. **INTRODUCTION**

The last decade witnessed a huge leap in the design and implementation of unmanned aerial vehicles and their networks. Governments, businesses, and smart cities deploy UAVNs in a variety of applications, such as emergency (search and rescue, disaster management, humanitarian, aid, ambulances), defense and security (traffic surveillance, drug monitoring, pipeline patrol, port security), environment (soil moisture, gas detection, agriculture), infrastructure monitoring and inspection (real estate, power line inspection, logistics, insurances), earth observation (archeology, GIS, media business), etc. With the evolution of the Internet of Things (IoT), cloud computing, advanced telecommunication systems, and artificial intelligence, UAVs are turning into smarter and more controllable machines. The Internet of Drones (IoD), in which the UAVs connect to the Internet either directly (for example, by connecting to a cellular base station) or via a GCS, is developing as one of the important use cases of UAVs.

With IoD becoming prevalent, many issues related to its implementation need to be considered. These include the intra- and inter-IoD communication methods and standards, air traffic management, data storage, energy resources, UAV physical security, data and network security, etc. In particular, IoD applications generate and use data that need to be secured against cyberattacks. Allowing attackers to steal users' sensitive information or forge wrong data leads to the failure of these applications. Several previous works illustrated that security-related issues in the IoD can be mitigated or resolved with the aid of powerful systems such as the blockchain (BC) [1–18]. The latter, being a distributed immutable database, protects the IoD data using cryptography mechanisms such as public-key encryption and hash functions. It is also used for guaranteeing the credibility of the stored data and for tracking the history of the UAVN operations.

The major problem that prevents the implementation of traditional BC models in the IoD is the limited resources of the UAV. The blockchain requires heavy computational and storage requirements for validating the data transactions, generating and storing the blocks, and attaining consensus. On the other hand, the UAVs are characterized by limited processing and storage capabilities due to the small size of the UAV. In addition, most UAVs utilize limited energy resources, such as lithium batteries, to fly and execute their mission. IoD administrators attempt to reduce the energy consumption of the drone to be able to extend its flight time. With the enormous amount of energy consumed by the underlying blockchain model and its consensus protocol, it becomes very difficult to integrate such model into the UAV while maintaining efficient processing and acceptable energy consumption.

Existing research projects that propose methods for integrating the BC into the IoD either limit the BC operations to the GCSs and/or cloud servers [3], [7], [11], [18] or apply a limited and lightweight BC within the UAVN [1–2]. Implementing the BC on the ground servers only leaves the drones susceptible to attacks and deprives them of benefiting from the data immutability advantage. Specifically, if the UAVs do not store a copy of the BC then their data will not be fully protected and could be stolen or modified by attackers. For this reason, several researchers considered deploying a lightweight BC model on drones. For example, Singh et al. [1] propose that each UAV will store a single BC block, and that the UAV can modify the body of the block by adding and updating transactions. However, such solution removes an essential characteristic of the BC, which is immutability. In addition, the system makes the drone store only the transactions in the last BC block and previous transactions will be stored on the GCS, which obliges the drone to send a request to the GCS each time it requires a previous transaction.

Another solution applied by some researchers is to reduce the size of the BC at the drone by applying lightweight techniques. For example, Ge et al. [2] propose using the Keccak hash function for signing the BC transactions, which reduces the overhead of the required cryptographic operations. In addition, the authors limit the size of the transaction data to 1KB in order to avoid storing large blocks at the drone. However, the system utilizes the Delegated Proof of Stake (DPoS) that requires the drones to engage in a voting mechanism to select the miners. Generally, DPoS is not designed for highly mobile and dynamic environments such as the IoD and it requires the drones that are selected as miners to spend energy in order to validate the new blocks. In addition, the technique proposed by the authors to limit the size of the transaction to 1KB in order to reduce the BC storage requirement is not reasonable since many IoD applications generate data that are larger than 1KB, which renders this system impracticable.

In this paper, we propose a fully lightweight and efficient blockchain model for the IoD. First, we classify the IoD data transactions into two categories: 1) that are needed by the drone and 2) that are needed by ground devices. We store each type of transaction into a separate type of BC block and make the drone save the blocks that contain the transactions it requires. We describe a mechanism for the drone to store part of the full BC in order to avoid overwhelming the drone's storage system with unneeded transactions. Second, we generalize our system by integrating the concepts of both public and private blockchain together within the proposed IoD blockchain. This is achieved by categorizing transactions into two main types: public and private; and applying public-key cryptography to private transactions. In addition, we provide a method for saving both types of transactions together within the block body. Third, we classify transactions based on their required security level. Since some transactions contain data that should be permanently secured, such as user-related data, smart contracts, etc., while other transactions contain data that should be secured during the UAVN deployment only (such as the drone's configuration and mission), it is a waste of resources to treat all transactions the same and apply the same strong cryptography operations to all of them. Hence, we apply different levels of security (encryption/signature) based on the security level of the transaction. This results in a very lightweight blockchain model at the drones. Finally, in order to take into consideration that many IoD applications are delay-sensitive, we modify the traditional consensus mechanism of the BC by introducing the concept of parallel mining. In our proposed consensus protocol, miners send requests to generate new blocks to a block orderer (BO) and the latter specifies to the miners the order of the blocks in the BC. This enables the miners to generate their blocks in parallel, leaving only the hash of the previous block. When the miner of the previous block obtains consensus on its block, the next miner can finish generating its block by calculating the missing hash, and so on. In the simulations, we show that this parallel multi-miner mechanism reduces the delay required to add the new block to the BC from 0.76s to 0.11s on average, which is a huge benefit to delay-sensitive IoD applications.

The remaining of this paper is organized as follows: in the next section, we provide an overview of the various models that apply blockchain technology in UAV systems. Section 3 contains the details of the proposed blockchain framework, while Section 4 analyzes the security aspects of the system. In Section 5 we test the proposed system by performing extensive simulations and comparing it to two BC systems for the IoD. Section 6 concludes the paper and illustrates several ideas for improving the proposed system.

## 2. LITERATURE REVIEW

The topic of integrating the BC into the IoD has been very hot in the last few years. Several research works proposed a variety of methods for achieving this integration. Some of these systems utilize the BC as a tool for protecting the data generated by the drones. Other models focus on the BC as a means for managing the IoD operations and storing the history of the missions performed by the UAVs for future tracking. The third type of system utilizes the BC to securely offer UAV services to cloud users by executing smart contracts and guaranteeing the safety of the users-drones interactions. In this paper, we utilize the BC for all three purposes and present several categories of BC transactions for each purpose.

As stated in the previous section, several IoD-BC systems limit the usage of the BC to the ground devices that interact with the UAVN, such as GCSs, cloud servers, and user devices. For example, Barka et al. [3] propose a mechanism in which each UAV that forwards a message adds a trust value to the message. The BC miners must decide whether to add a new transaction to the next block or not based on the trust ratings received from the UAVs. A similar system by Xu et al. [4] makes use of the UAVN to assist Mobile Edge Computing (MEC). The authors describe a framework for enabling resources trading between Edge Computing Servers (ECS) and UAVs. In the proposed system, the UAVs utilize the ECS resources in return for a certain incentive such as digital currency. The ECSs play the role of BC miners that implement Proof of Work (PoW).

A system that utilizes the BC to save the data of federated learning algorithms is proposed in [5]. Here, the drones assist the edge/cloud servers in executing the deep learning algorithm. UAVs participate in the mining process using PoW, which is unrealistic since drones cannot be equipped with the hardware required for PoW. Luo et al. [6] describe a system in which drones are used to cache data from IoT devices and forward the data to a MEC network that is outside the reach of the IoT devices. This method increases the coverage of the MEC network and saves the cost of deploying new MEC servers. The MEC servers participate in a private BC network. This results in developing a non-modifiable audit trail about which MEC server has been involved in processing each user's data. The authors of [7] propose a BC model that allows IoD applications to update the block headers and transactions by adding new parameters as required by the specific application. For example, a parameter named *Policy Header* is added to the block header to denote the type of operations that can be done by each user or group on the block transactions. The system groups the drones' users into clusters, where each cluster has its own ledger. Nodes in each cluster elect a forger node to generate the blocks of the cluster. The election process is done between a set of master controllers based on game theory. The BC (or ledger) is saved on the user layer only.

Several research works propose BC-based systems for 5G- or 6G- UAV networks. Aloqaily et al. [8] present a decentralized service delivery model, labeled Drones-as-a-Service in which drones are deployed on-demand to provide the required QoS connection to users in locations where 5G base stations cannot provide the demanded QoS. A user who requests to consume the services of a certain drone is given access to the keys of the drone to allow the user and drone to communicate in a secure manner. A system for anonymous authorization in IoD was proposed in [9]. The system hides the identities of UAVs within the exchanged message enabling only the receiver to trace back the identity of the sender. In this system, UAVs are lightweight BC nodes that store the header part only of the BC blocks. In [10], the authors describe an algorithm that enables each drone to obtain a session key from the GCS before exchanging secure messages. A main control room is used to register drones and GCSs. Transactions are forwarded by GCSs to cloud servers which act as BC miners. The BC proposed in [10] is private: only GCSs and authorized users are able to access the BC at the cloud servers.

Several BC systems propose grouping the UAVs into clusters. For example, Nguyen et al. [11] propose that each IoD cluster will have a drone leader which is usually a powerful drone. Drones in each cluster connect to the leader only, while the leaders connect to each other and to edge servers. Leaders collect, filter, process, and forward important data to the edge servers. The BC in [11] uses the RAFT consensus protocol and is operated and maintained at the edge servers only. Contrary to [11], the BC system proposed by Tan et al. [12] makes use of the cluster heads (CH) as BC miners. The cluster members keep only the BC state, which comprises lists of the members in each cluster. The authors use a lightweight election algorithm to choose the CH that will generate the next block.

The authors in [14] describe a flight compliance mechanism for Drone Service Providers that has two main objectives: 1) predefining the flight path of each drone based on the required services, saving it within the BC, and making sure that the flight path of each drone does not include restricted or private areas, and 2) assuring that each drone follows its flight path in order to avoid collisions. A similar approach is used in [15] in which drone owners register their drones with a UAV central authority. The latter offers UAV-as-a-Service via virtualization. Customers can buy the UAV services they want to consume via the APIs provided by the central authority. The latter defines the specification of each service, creates a smart contract for the customer, and saves it within the BC. Liao et al. [16] propose the Proof-of-Security-Guarantee (PoSG) consensus protocol, which gives a higher mining probability to nodes with better security guarantees. In PoSG, a certain amount of resources must be guaranteed and each resource must meet a minimum threshold for the drone to participate in the election.

The concept of software defined networking (SDN) is integrated with the BC in [17]. In the proposed system, the SDN controller is distributed among the GCSs, and the controller data and operations are managed via the BC. All GCSs share and synchronize the UAV control data in the BC. Data from the application plane (i.e., UAV users' applications) and the data plane (drones) are saved as new transactions in the BC, and the routing scheduling and flow table calculation logic are stored in the blockchain as smart contracts.

Singh et al. [1] propose ODOB, in which an Aviation Authority (AA) is used to define the access rights and take security-related decisions. The approach is based on decoupling data from the BC. When generating the hash of the previous block, only the block header is used in the generation, not the data. This method makes the BC block amendable and enables adding new transactions to it when needed. The authors of [2] propose several mechanisms within the IoD BC model: first, they apply lightweight cryptography hashing (Keccak) to sign transactions and generate the Merkle hash. Second, they add to the BC block two parameters: Policy List and Reputation Tree, which contain the access rights and the reputation of each drone. In order to reduce the size of the drone's BC, the system limits the size of each transaction to 1KB. In general, this assumption is not practical since many applications require the drone to send data that is larger than 1KB (such as images). This will result in dividing the data into many transactions and increases the communications and the block overhead size.

A zone-based architecture that uses a custom consensus method, labeled as the Drone-based Delegated Proof of Stake (DDPOS), was proposed in [18]. In this system, the smart city is divided into zones, and each zone contains many drones and a single drone controller (DC). The BC is stored and managed by the drone controllers. Each DC validates all the transactions inside its zone and approves the drones' truth during their movements in the smart city.

# 3. EFFICIENT LIGHTWEIGHT BLOCKCHAIN SYSTEM FOR THE IOD

## 3.1 System Architecture

The network architecture of our proposed IoD framework can be summarized as follows: we assume the existence of multiple UAV networks that can connect to each other via ground control stations (GCS). Each UAV network (UAVN) can connect to one or more GCSs. However, we don't assume a fixed architecture for all UAVNs. Rather, we consider that some UAVNs could be homogeneous; i.e., containing a certain number of the same drone that fly together in a swarm to perform a certain mission. Other UAVs may contain two or more types of drones where each type is assigned a specific task. For example, one type of drone could be equipped with environment monitoring sensors and lightweight processing and storage devices, another type could be equipped with a long-range communication module and configured to fly around in such a way to maintain the connectivity between the other drones and the GCS.

In general, different types of IoD applications will require different network configurations and types of drones, as pointed out in [13]. In addition, we don't assume any kind of clustering or zone dividing of the network, as in some previous works [11–12]. Since, in general, UAVNs will be used to execute different types of applications; such as emergency, environment and infrastructure monitoring, earth observation, etc. Each IoD application will require a different UAVN configuration and type of drones to execute it. Hence, no fixed network architecture and configuration will be suitable for all applications. However, we assume a general IoD architecture where the network can contain multiple UAVNs, with each UAVN having its own configuration, size, node types, and missions. Each UAVN can connect to the Internet via one or more GCSs, depending on the UAVN location and size. For example, a large UAVN deployed in an urban area could have multiple GCSs within the communication range of its drones; on the other hand, a UAVN deployed in a rural area could connect to the Internet via a single mobile ground gateway.

In addition, each group of GCSs will be controlled by a control authority (which could be physically centralized or distributed). For example, some GCSs will belong to a consortium of cloud service providers who cooperate to offer drone services to customers. Another group of GCSs could belong to a media company that deploys its UAVNs in news missions and events. A third set of GCSs could belong to a government agency that uses them in several smart-city applications such as road traffic management, pollution monitoring, healthcare, etc. In all these scenarios, each institution or consortium will use a central authority to manage and maintain a network of GCSs (a GCS could be physically fixed or mobile) that connect the institution's UAVNs to the Internet. In all cases, each drone in each UAVN should maintain a connection to at least one GCS, either directly or virtually via one or more other drones. The issue of connection management is related to the network routing protocol that will be used within the UAVN which is outside the scope of this paper. Several recent routing protocols for ad hoc networks, such as [19] and [20], proved their ability to efficiently connect mobile nodes to fixed infrastructure. Finally, the BC network will comprise the UAVs and GCSs which cooperate to generate and store the BC blocks. The control authorities will cooperate to monitor the IoD and blockchain operations in order to ensure their correctness and legitimacy and to take control actions whenever needed, such as resetting a GCS or drone when the latter is compromised by an attacker, or configuring the GCSs to add a new UAVN to the network. Each control authority (CA) will manage its network operations and monitor its blockchain data. In addition, multiple CAs can collaborate to take mutual decisions and perform common actions.

We consider a customized blockchain model for securing the IoD and saving its various operations as BC transactions. In the proposed system, we assume that certain types of data will be stored within BC blocks that are saved at the drone and GCSs, while other categories of data will be saved within blocks that need to be stored only at the GCSs. Usually, during a UAVN deployment, each drone will be given a specific task. Hence, each drone will continuously receive dedicated commands and send data related to its mission (such as GPS coordinates, speed, direction, power, sensor readings, images, videos, etc.) to the GCS. The configuration of the drone and the commands related to its mission should be sent in a secure manner from the GCS to the drone and stored securely in the partial copy of the BC within the drone storage system. On the other hand, all such data in addition to other data such as the information sent by the drones to the GCSs, the details of the execution of the drones missions, and the users' commands and consumed services are saved as different types of transactions within the full copy of the BC that is stored at the GCSs.

Taking into consideration that the BC could grow indefinitely and that the storage capabilities of the GCSs can be limited, it is important to differentiate between two main types of transactions in the BC. The first type will contain the complete data related to the transaction. This type is used whenever the transaction data has a small size or whenever the transaction should be stored on the drone. The second type will contain the transaction metadata and a reference to the actual data that will be stored on an external storage system such as a cloud data center or a big data cluster. The first type usually contains the drone's configuration, commands, mission details and updates, users' interactions, etc. while the second type will mostly comprise the drones' sensors' readings and data. Several previous works [2], [11], [18] assumed that all data will be saved within the BC, while other systems assumed that BC data should be saved on an external storage [10], [14]. Due to the fact that each UAV needs to get fast access to its configuration and mission data, it becomes vital to save this data securely within the UAV's BC. On the other hand, since many IoD applications require UAV nodes to send continuous streams of data to the GCS, it is necessary to use an external storage system to save this data while securing access to it via the BC.

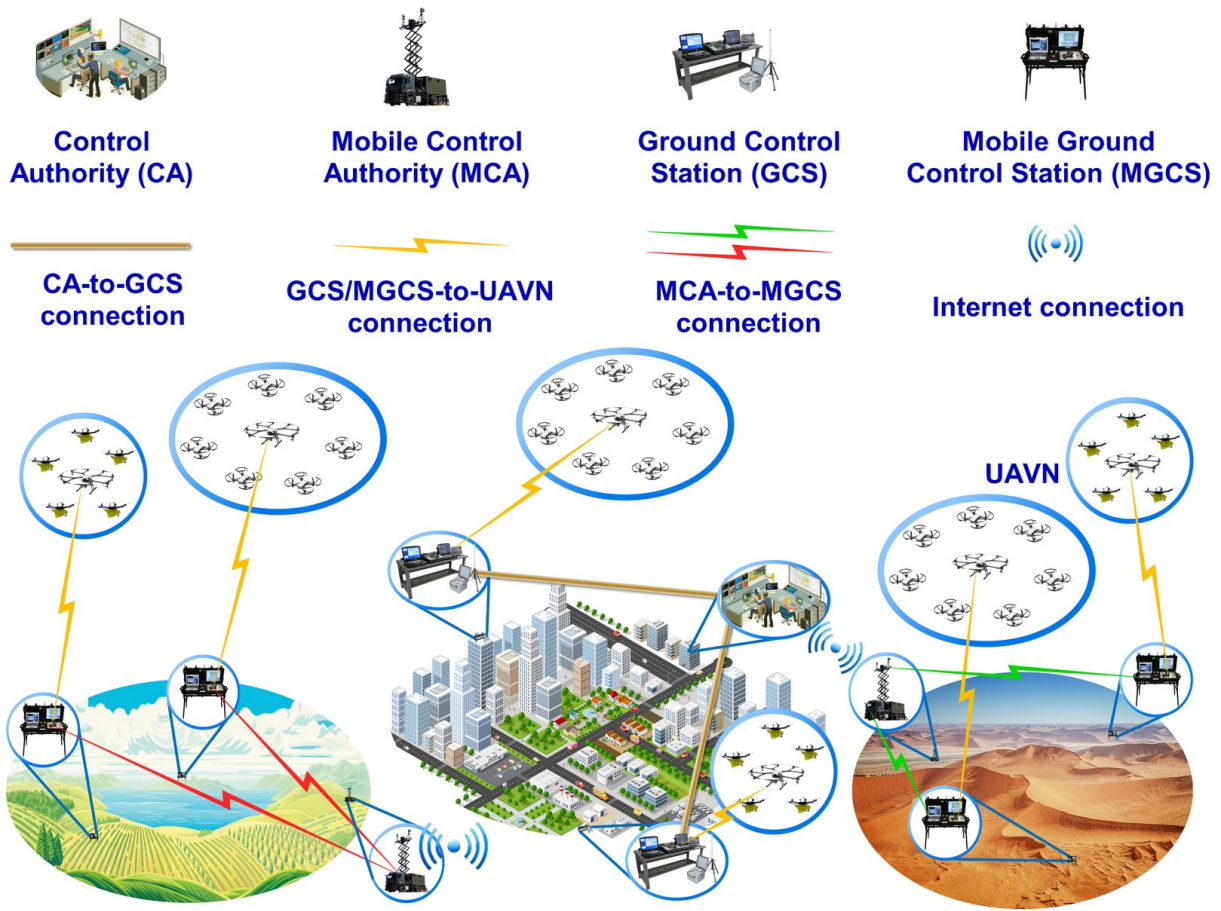

**Figure 1.** Components and connections of the proposed BC system

Note that in our model, we define a GCS as a control center that plays the role of connecting the UAVNs to one or more of the following entities: CA, cloud servers, data centers, IoD customers, and distant UAVNs. The GCS is responsible for deploying UAVNs and drones, assigning and updating the tasks of each drone, connecting administrators and customers to the drones (via dedicated APIs), receiving data from drones, participating in generating and storing new blocks (explained later), and reporting to the corresponding CA. A GCS can be a simple server mounted on a mobile vehicle (for example, a bus, truck, boat, etc.) that follows the UAVN and maintains its connectivity to the entities that were mentioned above (via WiFi or 5G), or could be a sophisticated control station that contains multiple servers and storage systems. Each CA will be responsible for establishing its GCSs and UAVNs, registering them within the BC system, and monitoring their operations and performance. An overview of the proposed system architecture and main elements is illustrated in Figure 1.

## 3.2 Threat Model

We assume that the control authorities are fully secured (both physically and digitally) and continuously monitored by security administrators to detect and handle attack attempts. On the other hand, we assume that GCSs are physically secured, but more vulnerable to cyberattacks than CAs. This is due to the fact that GCSs exchange information with UAVNs via wireless channels which could enable attackers to perform malicious activities. The drones are the weakest types of nodes in the system since they can be attacked both physically and digitally. Based on the taxonomy of attacks on UAVNs that was presented in [13], we outline in this section the types of attacks that we want to handle in our proposed mechanism. We briefly describe each of these attacks here. For more details on each attack, the reader can refer to [13]:

- Unauthorized access: occurs when sensitive, protected, or confidential data is accessed or modified by an unauthorized party. For example, consider the case when a UAVN is deployed for a search and rescue mission after a natural disaster, such as an earthquake. A drone that is equipped with thermographic sensors analyzes the sensors' reading in order to detect the presence of a trapped or injured person. When detecting such a case, the drone sends the exact location of the person to the GCS to help the rescue team locate the person as fast as possible. If an attacker is able to access and modify the data at the GCS for sabotage purposes, he could make the rescue team go to a false location and hence the life of the trapped/injured person would become in danger.
- Message forgery: in which a message is sent in such a way to deceive the receiver (GCS or UAV) about the identity of the sender. Continuing with the example from the previous point, an attacker could send a message from his/her device to one of the drones, as if it was sent from the GCS, ordering the drone to return to its base. This is another example of an attack that is performed for sabotage purposes. The attacker sends the packet while impersonating the GCS ID. If the drone is not configured with a security method to validate the ID of the sender, the attack will succeed and the drone cancels its mission and returns to its starting point.

- Malicious intruder: a UAV enters into the network silently and tries to forge other UAV credentials. For example, an attacker could use his/her own drone to infiltrate the UAVN and steal private data. In the described example, the attacker could fly his/her drone to the area affected by the earthquake and send Hello beacons to the nearby drones to discover their IDs and locations. Note that Hello beacons are usually used in wireless networks for packet routing purposes. After the attacker discovers the IDs of the nearby drones, he/she could impersonate one of the legitimate drones and send packets to the GCS using the legitimate drone's ID. The GCS will assume that the packets are coming from the legitimate drone and process them normally. This enables the attacker to attack the GCS. For example, the attacker can send false data about an injured person, or even attempt to trick the GCS to execute malicious code (malware).

- Sybil attack: a malicious node uses multiple identities at the same time to gain higher influence in the network and perform illegal actions. In our example, suppose that each drone that correctly sends the location of an injured/trapped person to the GCS receives an incentive (for example, a small amount of digital currency). An attacker can register with the CA using multiple identities and use each of these multiple identities to send several packets to the GCS and collect the incentive multiple times.

- Eavesdropping: an attacker secretly listens to the communication between two nodes (UAVs and/or GCSs). Going back to the example in which the attacker infiltrates the UAVN by flying his/her drone to the area affected by the earthquake, the attacker can position his/her drone between two drones that are exchanging packets and save a copy of each packet. If the drones are exchanging private or sensitive data, the attacker will be able to illegally acquire a copy of that data.

- Man in the middle: an attacker monitors and modifies the messages exchanged between two nodes. This attack is similar to the previous one (i.e., eavesdropping), with the difference that in eavesdropping the attacker just listens to the exchanged packets (passive attack), while in "man in the middle" the attacker inserts him/herself in the middle of the communication and modifies the packet that he/she receives before sending it to the intended destination (active attack). For example, an attacker who positions his/her drone between two drones that are exchanging packets, can change the data in the packet and then forward it to the destination drone. The destination drone will receive packets that contain malicious data. The normal operations of the system will be affected when the destination drone processes the malicious data. For example, suppose that a drone $D_1$ is sending a packet that contains the location of an injured person to a drone $D_2$. $D_2$ has direct communication with the GCS and will forward the packet to the GCS. An attacker who intercepts the packet before it reaches $D_2$ could send a jamming signal to prevent the packet to reach $D_2$, modify the location inside the packet, and then forward it to $D_2$. This will cause $D_2$ to send a wrong location to the GCS.

- 51% attack: an attacker or group of attackers attempt to control more than 50% of the mining hash rate in order to prevent certain transactions from gaining confirmations or achieve illegal benefits. For example, in the described search and rescue scenario, suppose that an attacker is able to compromise more than half of the trusted GCSs in the BC network (trusted GCSs, or TGCSs, will be used to mine the BC blocks, as we describe in Section 3.7). In this case the attacker will be able to corrupt the blockchain by inserting malicious transactions or even complete blocks. If each drone receives an incentive for sending a correct location of an injured/trapped person, then the attacker could modify all the transactions in the compromised GCSs by putting his/her ID as the sender of each transaction and illegally receives all the incentives. If the attacker compromises more than 50% of the miners, he/she will be able to generate consensus on his/her malicious transactions, add them to the blockchain, and have the incentives illegally transferred to the attacker's account.

### 3.3 Transaction and Block Types

In our proposed model, we define two types of blockchain blocks: the first type will be stored as part of the drone's BC and on the full BC at the GCSs, while the second type will be stored on the GCSs BC only. The first type of block (Block$_{T1}$) can contain one or more transactions that are related to the following types of information:

- Data related to the drone setup and configuration, such as the flying path coordinates (could be updated via several transactions that contain updates to the drone's predefined path), desired speed, flights height(s), changes to the yaw; roll; and pitch angles, etc.

- Commands sent by the GCS to the drone. These commands instruct the drone to activate an actuator, take an image, drop an object, send a message to another drone, land down, etc.
- Blockchain smart contracts that are triggered based on specific events and execute a list of instructions based on the agreement between the drone service provider and the cloud user. For example, the smart contract can be used by the user to analyze certain drone readings via machine learning algorithms to produce specific outputs.
- Security-related data that are sent by the drone to the GCS, such as reporting a malicious behavior or incident. In such case, the GCS forwards the transaction to the CA who will gather the data related to the incident from the drones and GCSs and take the necessary action.

The second type of block ($Block_{T2}$) is stored within the GCSs BC only and contains transactions that are related to the following information:

- Various types of data sent by the drone to the GCS, such as information related to the drone's mission (flight parameters such as geographic coordinates; fuel; battery power; airspeed; payload; etc.), sensors' readings, images, videos, results of executing a command, etc. When the GCS receives this kind of data it either adds the data to a new transaction or stores the data into the predefined storage system and adds the reference to the data to a new transaction, as we previously explained.
- Data related to cloud users who interact with the drones via virtual APIs. This type of data includes users' accounts and access rights, performed activities, used resources, etc. A cloud user connects to the cloud server of the drone service provider and then his/her connection is redirected to the GCS within which the user can utilize certain APIs (based on the user's access rights) to connect and interact with the drones.
- Data related to the CA commands and management of the GCS, such as control commands related to drones' configurations (add/modify/delete), drone data (select/modify storage location; change storage characteristics, etc.), user's related (allocate/modify resources; add/modify access rights; etc.), add new UAVN/drone, stop one or more drones and return them to base, block the certificate of one or more BC IDs, etc.

This division of the BC blocks is made in order to allow drones to store only the data that will be required and used by the drone during its flight, and store the remaining data on the GCS BC. Using this strategy will allow us to utilize the limited storage space at the drone. Previous works, such as [2], allowed the drone to store all types of transactions, but limited the size of each transaction to 1KB to keep the size of the block small. However, such assumption is not realistic since many types of transactions that we described in this section will contain data that is larger than 1KB. Hence, our solution for storing the BC within the limited storage space on the drone is to make the drone store only the blocks that contain the data that is necessary for the drone, in addition to applying dynamic and lightweight cryptographic mechanisms inside the BC, as we will explain in the next sections.

### 3.4 Classification of Transactions based on Users' Access

In addition to classifying each transaction based on whether it will be saved within the UAVN or not, we divide transactions into three types based on the access rights of the BC nodes. In our proposed BC framework, we assume that part of the data in the BC will be public (i.e., similar to a public BC), and the other part will be private (similar to a private BC). In turn, this last part will be divided into two subtypes: transactions that can be accessed only by a single owner, and transactions that can be accessed by a group of BC nodes. The detailed classification of transactions based on access rights is as follows:

- Type 1 ($Tran_{T1}$): Transactions that can be accessed by all nodes in the BC network. These transactions constitute the public part of our proposed BC model. These transactions will be stored within the body of the BC block without being encrypted by a cryptography key. However, each transaction will be signed by its creator, as usual. Examples of these transactions include: 1) adding a new UAV to a UAVN (this transaction contains the ID, starting position, public key, and general mission/role details of the new UAV), 2) modifying the mission/role of an existing UAV, 3) public data (or data reference) that was sent by a drone to the GCS, etc.

- Type 2 ($Tran_{T2}$): Transactions that can be accessed by one BC node. This node will be the transaction owner, which could be a drone or a GCS. This type of transaction is encrypted with the public key of the owner before saving it in the BC block. This enables only the owner and the CA to access the transaction in the BC (since the CA of each institution will store the keys of its GCSs and drones). The owner can retrieve the transaction to use its data, while the CA can access the transaction to track the history of an event, check a security incident, or ensure the validity of certain information. Examples of this type of transaction include:

1) Drone configuration parameters and commands that are sent by the GCS to the drone. Here, the GCS generates the transaction, encrypts it with the drone's public key, and adds it to a new BC block (based on the consensus algorithm that we explain in the next section). Note that such block will be of type $Block_{T1}$, i.e., will be stored at both the GCS and the drone. Hence, the drone can retrieve the transaction from the block and use its data.

2) Another example of this transaction is that containing user-related data. Here, the GCS will collect the data from the user's session (interactions with UAVN), encrypt the transaction with its public key, and add it to the next block in the GCSs BC (details on BC generation and storage are explained in the next section).

3) A third example is a transaction that contains the security-related data that is sent by the drone to the GCS. Here, the drone encrypts the data with the GCS public key and sends the transaction to the GCS. The latter adds the transaction to its next block, and then it decrypts the transaction and processes the data.

4) The last example of such transaction is that containing the CA commands to the GCS. Here, the CA encrypts this transaction with the GCS public key and sends it to the GCS. The latter will store the transaction in the GCSs BC and then decrypts it to execute the CA commands.

- Type 3 ($Tran_{T3}$): Transactions that can be accessed by a group of BC nodes. For these transactions, the CA assigns a group key pair (public/private keys) to the group of nodes that will access the transaction. Hence, the transaction will be encrypted with the group public key so that only the group members will be able to decrypt it. An example of this kind of transaction is a common task command that is sent by the GCS to a group of drones (i.e., all drones in the group will execute this task). In this case, the GCS encrypts such transaction with the group public key and saves it in a $Block_{T1}$ block that will be saved within the BC of the group members. Another example of such transaction is that containing a user's smart contract that will be executed by several drones (however, if the user's smart contract will be executed by a single drone then this transaction will be of type $Tran_{T2}$).

Based on this classification, we say that our blockchain model combines characteristics from both public and private blockchain systems in order to avoid encrypting transactions whose data is public, and hence, reduce the overhead on the drones; while at the same time securing the private transactions by encrypting them with the key of their owner or owners. Note that the CA will be responsible for generating and storing the history of key pairs for its GCSs, drones, and groups of drones. This is useful in case there is a need to decrypt a $Tran_{T2}$ or $Tran_{T3}$ transaction in the future (after the UAVN is dismantled).

### 3.5 Lightweight Blockchain Cryptography

In order to take into consideration that different types of data that are generated in an IoD require different levels of security, we design a model in which we vary the power of the cryptographic operations that are applied to a transaction according to the required security level of the transaction. First, we define two main types of transactions:

- Type 1 ($Tran_{S1}$): Permanent-security transaction. This type contains data that need to be permanently secured, such as user's smart contracts, user's access rights, user's session details, CA commands to the GCS, private data sent by the drone to the GCS, etc. This type of transaction will be signed and encrypted (for $Tran_{T2}$ and $Tran_{T3}$ transactions) using powerful cryptography algorithms (such as elliptic curve cryptography or ECC) to ensure that they are well secured.

- Type 2 ($Tran_{S2}$): Temporary-security transaction. This type contains data that should be secured for a temporary period (for example, the duration of the UAVN deployment). Examples of such types of transactions are those containing the IDs and specifications of the deployed drones, the configuration of a drone, the commands sent from the GCS to the drone, and the security-related data that are sent by the drone to the GCS. Such a transaction will be signed and encrypted (if it is of type $Tran_{T2}$ or $Tran_{T3}$) using lightweight cryptographic algorithms that will be customized by reducing their security power (for example, by reducing the size of the cryptographic key, the number of rounds in the algorithm, etc.) in such a way to ensure the security of the transaction for the required period (i.e., an attacker will not be able to deduce the keys that are used to decrypt/sign the transaction during the required period).

Using this lightweight cryptography mechanism for $Trans_{S2}$ transactions will reduce the processing and storage load on the drones which will enable the drone to store a larger number of blocks before its memory is full (more details on this issue in a later section), and at the same time reduce the processing delays required for generating, reading, and validating the transactions in the drone's BC. In Section 5, we prove this fact via simulations and illustrate that implementing our proposed model for lightweight blockchain cryptography reduces the processing delay at the drone by 86% and the transaction size by 83.5%, on average. Note that our proposed solution is general and doesn't impose any specific algorithm for signing/encrypting the transactions or any specific method for reducing the power of the utilized algorithms to reduce their overhead. However, we describe in Section 5 the method that we used in our simulations to perform these operations by using three versions of Elliptic Curve Cryptography (ECC) and two versions of the SPONGENT hash functions.

## 3.6 Blockchain Structure

In order for the mechanisms that we proposed in the previous sections to work, the blockchain structure used in our framework is designed as shown in Figure 2. In this model, two new parameters are added to the block header, the first is the *Block Type* (*BT*), while the second is the *Transaction Access* list (*TA*). The *Block Type* will be either $Block_{T1}$ or $Block_{T2}$, depending on whether the block will be saved by the drones or not (as explained in Section 3.3). The *Transaction Access* list contains the type of each transaction in the blockchain body, as described in Section 3.4 (i.e., $Tran_{T1}$, $Tran_{T2}$, or $Tran_{T3}$). If the transaction is of type $Tran_{T2}$ or $Tran_{T3}$, the list will contain the ID (or IDs) of the transaction owner (or owners). Figure 2 shows an example of a block that contains transactions of the three types. Transaction $T_1$ is of type $Tran_{T1}$ and hence it is public. On the other hand, $T_2$ is of type $Tran_{T2}$ and its owner is the node whose ID is $N_{21}$. Hence, $T_2$ will be encrypted with the public key of $N_{21}$. Transaction $T_3$ is of type $Tran_{T3}$ and its owners are nodes $N_{21}$, $N_{33}$, and $N_{37}$. Hence, $T_3$ will be encrypted with the group public key that is assigned by the CA to the three owners.

Note that if a node attempts to access a transaction that it doesn't have access right to, the drone's blockchain system will detect that from the *Transaction Access* list, generate a security-incident transaction, and send it to the GCS after encrypting it with the latter's public key. Also, note that a node $N^*$ who is requesting access to a transaction that it has access to will generate its signature and add it to the request message in order for the receiving node to authenticate $N^*$ and give it access to the transaction. This makes an attacker unable to impersonate a node unless it steals its private key.

The second change to the traditional BC structure is adding new parameters to each transaction in the BC body. These parameters include: *Security Level* (*SL*), which is the security type of the transaction ($Trans_{S1}$ or $Trans_{S2}$); $Enc_{ID}$, which is the ID of the cryptography algorithm that was used to encrypt the transaction; $Hash_{ID}$, which is the ID of the hash function that was used to sign the transaction; $Enc_{par}$, which contains the parameters of the cryptography algorithm that was used to encrypt the transaction, and $Hash_{par}$, which contains the parameters of the hash function that was used to sign the transaction. Note that the values of $Enc_{ID}$ and $Enc_{par}$ will be equal to NULL if the transaction is of type $Tran_{T1}$. These parameters will be merged with the encrypted transaction. These parameters are essential since each transaction may be encrypted and/or signed with different parameters as explained in the previous section. If the transaction is of type $Tran_{T1}$, then the reader will use the $Hash_{ID}$ and $Hash_{par}$ to know the hash function and the parameters that should be used to decrypt the signature and validate the transaction. If the transaction is of type $Tran_{T2}$ or $Tran_{T3}$, then the reader will use the $Enc_{ID}$ and $Enc_{par}$ to know the algorithm and the parameters that should be used to decrypt the transaction and then uses $Hash_{ID}$ and $Hash_{par}$ to decrypt the signature and validate the transaction. The individual parameters within $Enc_{par}$ and $Hash_{par}$ vary depending on the utilized algorithms and are written using the notation (name=value). For example, key_ID=4, Nb_rounds=15, etc. Here, key_ID identifies the ID of the key that was used to encrypt the transaction (different keys with different sizes), and Nb_rounds denotes the number of rounds that were executed during the encryption process.

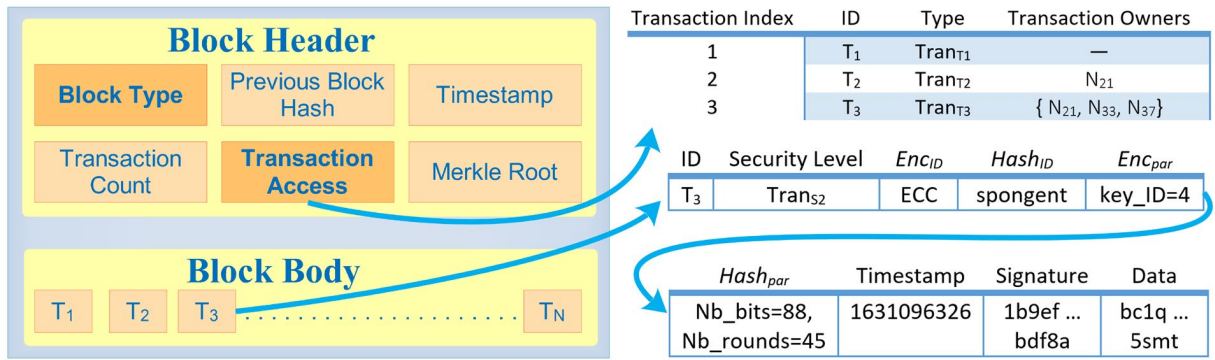

**Figure 2.** Block structure of the proposed BC model

## 3.7 Block Generation and Consensus

When utilizing a blockchain system to secure a network, the security of the network becomes largely dependent on the security of the blockchain miners. The miners are responsible for guaranteeing the correctness and legitimacy of all exchanged transactions, and controlling access to the BC. By guaranteeing the authenticity and trustworthiness of BC miners, the stored transactions are secured. The main method that is used in BC networks to ensure that BC miners are working legitimately is the consensus protocol, which is applied by all BC miners to reach an overall agreement on the validity of each transaction before it can be added to the BC.

Traditional blockchain consensus models, such as PoW, PoS, and DPoS are not suitable for networks that use the BC for saving the history of data transactions. The main reason is that the aforementioned protocols were designed such that BC miners are ensured to be trustworthy and legitimate either by making them perform extensive computations to earn an incentive (PoW), or by checking their high share in the blockchain network (PoS and DPoS). These methods cannot be applied when the blockchain is used for saving data transactions instead of financial ones (i.e., when the BC is not intended as a distributed ledger). For example, GCSs in the IoD are not expected to be equipped with mining ASICs since their purpose is to manage the UAVNs not to engage in a mining competition.

For these reasons, consensus algorithms that are used in a data blockchain system need to find a different method for ensuring the reliability and legitimacy of the nodes that will generate and validate the BC blocks. In a previous paper, we proposed the Proof of Accumulated Trust (PoAT) protocol for the Internet of Vehicles (IoV) [21]. In PoAT, we select the miners among roadside units (RSUs) based on their accumulated trust. We briefly explain the concept of the accumulated trust here: for an RSU to be selected as a Trusted Node (TN) and become a BC miner, it needs to satisfy two conditions:

1) The RSU should accumulate a total number of trust points greater than a threshold $TH_{TN}$ during a period $T_{TN}$.
2) The RSU should accumulate a number of trust points greater than a threshold $TH_m$ during each subperiod $m$ of $T_{TN}$ (to ensure the legitimacy of the RSU in each subperiod).

The details on how to define the values of $T_{TN}$, $TH_{TH}$, and $m$ were discussed in [21]. Also, note that an RSU gains trust points by participating in legitimate operations and loses trust points by engaging in malicious operations. The complete details on the PoAT consensus algorithm can be found in [21].

In this paper, we adopt a consensus model similar to PoAT. In the proposed method, we select a set of miners among the GCSs based on their accumulated trusts. We will call each member in this set a trusted ground control station or TGCS. The remaining GCSs will save the BC without acting as miners. Also, the drones will not act as miners due to two reasons: 1) limited computational and energy resources, and 2) high vulnerability to attacks. However, each drone will store part of the blockchain within its system (more details on this later), and will be able to access the remaining part via the GCS (since we assumed a virtual connection should always exist between each drone and at least a single GCS).

In general, the IoD shares a lot of common characteristics with the IoV. For example, both networks contain two main types of nodes: moving vehicles (cars, drones) and fixed infrastructure (RSUs, GCSs). In both networks, the vehicles move at a high speed, have intermittent connectivity with other vehicles, and connect to the fixed infrastructure in order to reach other nodes in the network or to access online services and data [22]. In addition, drones usually move in fixed flight paths that contain waypoints, similar to cars' movements on roads that contain intersections. The major difference between cars and drones is that the former can be equipped with much more resources (processing, storage, and network) than the latter. This is the main reason why a lightweight mechanism is necessary for the IoD. In our system, we adopt the consensus protocol that was described in [21] based on the fact that GCSs resemble RSUs to a high degree. Hence, TGCSs can be selected in a very similar approach to that used in [21] to select TNs. In addition, the cars were not chosen in [21] to play the role of BC nodes, since their security cannot be guaranteed. Similarly, in this paper, we do not allow the drones to participate in the blockchain operations, mainly validation and consensus, due to the same reason and also because of the drone's limited resources. The described BC operations are performed within the network of GCSs and TGCSs, which resembles the network of RSUs and TNs in [21].

Due to the fact that many IoD applications are very delay-sensitive, we modify the PoAT protocol [21] in order to reduce the delay between the time at which the data is generated and the time at which it is added to a new BC block. In traditional consensus algorithms, such as PoW and PoS, miners compete to add their blocks to the BC. The miner that wins has its block added, while other miners should wait for the next round. In IoD networks, drones in a UAVN send their data to the GCS. Hence, each GCS will have its own data. It is not possible to have all GCSs exchange all their data transactions to ensure that the next block will include all data that was generated by all drones in all UAVNs. This is due to the fact that many IoD applications produce data all the time, which means that the number of data transactions will be enormous and cannot be broadcasted altogether to all GCSs or else the network will be overwhelmed with transaction packets. On the other hand, making each GCS generate its own block means that each GCS needs to wait its turn to add its block to the BC, which delays the applications that need to process this data. Hence, the traditional method of generating a single block by a single miner at a certain instance is not suitable in large networks in which nodes generate a huge amount of data that should be added to the BC within a certain time limit (after it is generated) in order to be efficiently processed by delay-sensitive applications. For example, in emergency or disaster relief scenarios, data generated by a drone that contains the location of a wounded person should be added to the BC within a very small delay (for example, less than 1 second) so that the rescue team can head towards the location of the wounded person before it is too late. Hence, we need to find a way in which each TGCS will have its block added to the BC as fast as possible.

We propose a Parallel Multi-Miner Proof of Accumulated Trust (PROACT) consensus protocol for the IoD. In PROACT, at the start of the system, each CA selects some of its GCCs to act as TGCSs. The selected GCSs must be highly secured to reduce the risk of their compromisation. Since TGCSs will act as the BC miners, it is important that each TGCS is highly secured using advanced information security mechanisms and controls. Since the IoD is dynamic, one or more TGCSs might be shutdown or replaced at a later stage. In addition, a TGCS must be replaced and restarted if it is compromised. In such cases, a new TGCS must be selected in order to maintain a sufficient number of BC miners in the network. The process of selecting a new TGCS was described in [21] (Section 3.2. The accumulated trust mechanism). We summarize this process here briefly: When a new TGCS is to be selected, the CA chooses from its TGCSs the one that is most secured and that satisfies the two trust conditions that were mentioned in the second paragraph of this section. Using this strategy, the most secured GCSs that proved their trustworthiness during the network lifetime are selected as TGCSs.

Each CA should select a number of TGCSs between 1 and $TH_{CA}$, where $TH_{CA}$ is the number of TGCSs per CA and can be calculated as $TH_{CA} = Max_{TN}/N_{CA}$, where $Max_{TN}$ is the maximum number of TGCSs that could exist in the network (details of $Max_{TN}$ are explained in [21]), and $N_{CA}$ is the total number of CAs. After each CA selects its TGCSs, an algorithm will be used to randomly assign a certain number of GCSs that are not TGCSs to each TGCS. In other words, each TGCS will be assigned a number $N_{GCS}^{TGCS}$ of GCSs that are not TGCSs and that are selected randomly. When the number of TGCSs in the network changes (for example, when a CA decides to add a new TGCS or stop a current TGCS), the algorithm is re-executed and each TGCS is re-assigned a new number of random GCSs. Suppose that at a certain instance in time the total number of GCSs (that are not TGCSs) is $N_{GCS}$ and the total number of TGCSs is $N_{TGCS}$, then each TGCS will be assigned a number of GCSs $N_{GCS}^{TGCS} = N_{GCS}/N_{TGCS}$ (which will be rounded randomly to the lower or higher integer, while the last TGCS will be assigned the remaining GCSs).

All TGCSs will be generating their blocks in parallel. However, in order to define how will the blocks be ordered in the BC (which block will come after the other), one of the CAs will play the role of the Block Orderer (BO). The role of the BO will be exchanged between CAs, i.e., it will be assigned to a new CA after a certain period. When a TGCS wants to start generating a new block, it sends a New Block Request (NBR) message to the BO and then starts generating the body of the block by ordering the transactions and generating the parameters in the block header (except the hash of the previous block). The NBR will contain the timestamp at which the TGCS sent it. The BO will receive NBRs from the TGCSs and order the blocks based on the timestamps in the NBRs. For example, suppose that the last block that was added to the blockchain had a sequence number (or ID) equal to 43. After that, the BO receives three successive NBRs from three TGCSs: first, it receives an NBR from $TGCS_2$ with timestamp $T_2$, then it receives an NBR from $TGCS_1$ with timestamp $T_1$, and then it receives an NBR from $TGCS_3$ with timestamp $T_3$. Now, suppose that $T_1<T_2<T_3$ (i.e., $T_1$ oldest, $T_3$ newest). Hence, the BO assigns to the new block of $TGCS_1$ the ID 44, to that of $TGCS_2$ the ID 45, and to the new block of $TGCS_3$ the ID 45. Next, the BO broadcasts to the TGCS network the IDs of the new blocks and their TGCSs. Hence, when $TGCS_1$, $TGCS_2$, and $TGCS_3$ receive the broadcast message, each of them will know the ID of its new block (i.e., its order in the BC). Next, $TGCS_1$ will finalize its block by hashing the block whose ID is 43 (which is the last block in the BC), and sends it to the TGCSs to validate it. After the block of $TGCS_1$ is validated and added to the BC, $TGCS_2$ will directly hash it, add its hash to its block, and send it to TGCSs to validate it, and similarly for $TGCS_3$. Using this method, a TGCS does not need to wait for the previous block to be generated and added to the BC before it starts generating its own block. Rather, each TGCS will work in parallel with other TGCSs in generating its new block (except for the hash of the previous block) and then will wait its turn (according to the order that was specified by the BO) to add its block to the BC. This strategy will reduce the time needed to generate new blocks in general since miners will be producing their blocks in parallel. Figure 3 illustrates the various operations within PROACT.

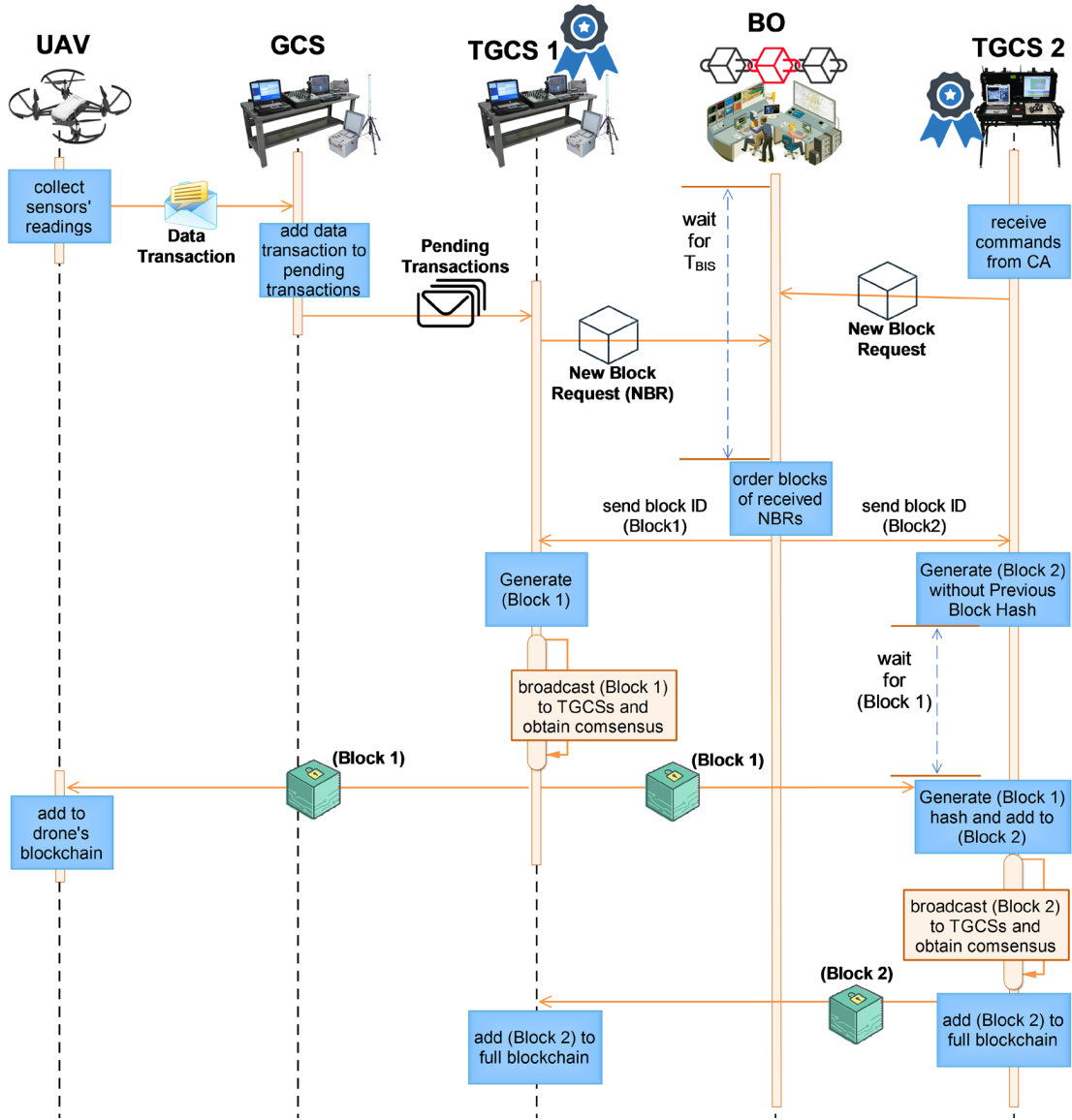

**Figure 3.** Various operations made by the GCSs, TGCSs, and BO when executing the PROACT protocol

Each GCS will send its transactions to the assigned TGCS. The TGCS will check the validity of each transaction, order the transactions in the body of the block, generate the new block (except the hash of the previous block), and then wait for the previous block to be added to the BC. When the previous block is added, the TGCS generates its hash, completes generating its block, and then broadcasts it to the other TGCS to validate it and add it to the blockchain based on the consensus mechanism that was proposed in PoAT (i.e., via "Block ACK" and "Block ERROR" messages; complete details in [21]). This operation will be executed in parallel by all TGCS. Hence, each TGCS will be doing the operations of generating, broadcasting, and obtaining consensus on its block, and validating the blocks of other TGCSs in parallel. In addition, the TGCS can generate multiple new blocks if it needs so (for example, if the TGCS accumulates enough new transactions to start a new block before it adds its waiting block). Note that the TGCS will use a dedicated network channel for sending its own blocks and their related messages (such as NBR), and another network channel for receiving the blocks of other TGCSs to validate them.

The block generation and consensus methods applied in PROACT reduce the number of transactions per block as compared to traditional consensus in which a single miner generates each new block. In PROACT, we divide the total transactions between all the TGCSs (initially, each transaction will be at the GCS that generated it or received it from a drone, then each GCS will send its transactions to the TGCS to which it is assigned). Each set of transactions at a TGCS will be processed in parallel with the other sets at different TGCSs. Hence, the time needed to validate and add new blocks to the BC will be much reduced, as we will see in Section 5. Consequently, each UAV will be able to add new blocks to its BC and make use of the data in these blocks in a much faster manner. In addition, the administrators and cloud users will be able to access the drones' data from the BC much faster.

Note that in many cases the new block might contain a single transaction (such as UAV blocks). In other cases, the block will contain many transactions. In all cases, the transactions that are usually saved in $Block_{T1}$ blocks will be saved in separate blocks and stored in both the UAVs and the GCSs BCs, while the transactions dedicated to GCSs (i.e., those that are usually saved in $Block_{T2}$ blocks) will be saved in separate blocks that are stored within the GCSs BCs only. Also, note that a TGCS could be generating a $Block_{T1}$, $Block_{T2}$, or both at the same time. If the TGCS wants to generate two blocks at the same time, it will send to the BO an NBR that contains two requests for the two blocks.

Each time a GCS adds a new block of type $Block_{T1}$ to its copy of the full BC, it examines the *Transaction Access* list in the header of the block and identifies the drones within its UAVNs that own one or more transactions in this block (i.e., they are listed as owners for one or more transactions). For each such drone, the GCS sends a copy of the block to the drone. Hence, each drone will save a part of the full BC. The BC of each drone will contain the blocks that contain the transactions that it owns. In this way the drone will have fast access to its transactions. If the drone requires a transaction that doesn't exist in its BC, it will request it from the GCS or from a neighboring drone (if the requested transaction is owned by a neighboring drone).

Note that the BO will assign new IDs to new blocks as follows: When the BO starts, it generates the *Genesis* block which contains the IDs and public keys of all GCSs and the IDs of the TGCS (the initial selection of the trusted nodes was described in [21]). The BO will obtain the information that it will add to the *Genesis* block from the CAs. Next, the BO broadcasts the *Genesis* block to the GCSs. After that, the BO will wait for a certain small time $T_{BIS}$ (BIS = Block ID Assignment). If it receives one or more NBRs during $T_{BIS}$ then it orders them based on the timestamp and sends the IDs of the new blocks to the TGCSs as explained before, and then starts a new $T_{BIS}$. At the end of each $T_{BIS}$, the BO repeats the same operation (i.e., examining the NBRs that it received, ordering based on timestamps, assigning block IDs, and broadcasting new assignments to TGCSs). Note that a certain NBR might be delayed to reach the BO in some cases, and it arrives in the next $T_{BIS}$ after it was sent, which makes other NBRs that have higher timestamps obtain smaller IDs. However, this delayed NBR that arrived in the next $T_{BIS}$ will get the first ID in the next $T_{BIS}$.

## 3.8 Drones' Blockchain

A major distinction between our IoD blockchain model and previous ones is that we provided a method for each drone to save part of the full IoD BC. As stated in the previous section, each TGCS will select from the transactions those that are owned by the drone (such as the drone configurations and mission commands) or contain data that should be processed by the drone (such as users' smart contracts) and will save these transactions in separate $Block_{T1}$ blocks and send them to the drone. In addition, transactions that are owned by several drones are saved in separate $Block_{T1}$ blocks and sent to all their owners. Hence, the proposed framework enables the drones to store in its blockchain the blocks that contain data that are important to it only. In addition, the lightweight cryptography mechanisms will enable the drone to store and process these blocks with limited storage and processing capabilities.

Taking into consideration that the memory and storage devices within the drone will have limited capabilities, the size allocated to the drone's BC will be limited. When the drone wants to save a new block and it finds that the size of its BC has reached its limit (the limit will be identified based on the drone's total storage capability), it will execute a Block Replacement Algorithm (BRA) that selects one or more of the blocks in its BC and deletes them in order to free enough space to save the new block. The BRA will follow a replacement strategy that will be defined by the UANV administrator, such as the oldest block in the drone's BC, the block that contains outdated data (for example, data that has been updated in newer blocks), etc. In all cases, the drone deletes the block(s) that were selected by the BRA and replaces them with the new block(s). If the drone needs to access data from previous blocks that were deleted, it sends a BC request to the GCS. Note that blocks in the drone's BC are ordered based on their IDs (or sequence numbers) for searching purposes. However, the drone's BC is not used for validating the correctness of blocks, since it will contain a limited number of the blocks from the full BC at the GCSs. Rather, the process of block validation will be done at the GCSs whenever needed. The UAVN administrators should ensure that the drone contains sufficient storage space to save all the important BC blocks that it will need during its mission.

## 4. SECURITY ANALYSIS

In this section we describe and analyze the techniques that are used by the proposed system to defend against the various attacks that were described in the threat model (Section 3.2):

- Unauthorized access: the proposed BC model added a new parameter to each block, which is the *Transaction Access* list (*TA*). The *TA* is used to enforce access control policies on each transaction in the BC. In addition, the proposed system adopts a mixture of public-private BC models, in which public transactions are stored without encryption, while private transactions are encrypted with the owner's public key in order to prevent unauthorized access to private data. Recalling the search and rescue scenario in Section 3.2, the attacker will not be able to access a private transaction at the GCS since the attacker's ID will not exist in the *TA* entry of that transaction.

- Message forgery: each transaction that is sent by a drone or a GCS is signed with the owner's digital signature. The transaction is stored in the BC along with the signature, which validates the identity of the transaction generator. If a malicious node tries to send a malicious transaction by impersonating a victim node, it will not be able to sign the transaction since it doesn't have the victim's private key. In the search and rescue scenario, if an attacker sends a forged message impersonating the GCS to one of the drones, the latter will detect that the message is forged by checking the signature of the transaction. Since the attacker doesn't have the GCS private key, he/she will not be able to sign the transactions correctly.

- Malicious intruder: similar to the previous point, a malicious UAV will not be able to send malicious transactions or query the BC by impersonating a legitimate node, unless it has the victim's private key (to generate the signature). Usually, the drone's private key is securely saved within hardware tamper-proof devices. Hence, the attacker needs to attack the drone physically and crack its system to be able to steal the private key and use it. However, such operation will be detected by the GCS (if the drone becomes offline for a long period) and the drone's keys are invalidated. This prevents the attacker from attacking and impersonating a legitimate node. In addition, an attacker cannot use an invalid drone ID, since the other drones will detect that this ID does not exist in the *Genesis* block (explained in Section 3.7), and will isolate the attacker.

- Sybil attack: similar to the previous point, a malicious drone will not be able to use the identity of another drone successfully, unless it steals the private key of the victim. In addition, an attacker will not be able to assign multiple IDs to his/her drone, since each drone ID will be linked with the real ID of its owner (such as the SSN).

- Eavesdropping: in order to thwart this attack, all private or sensitive data will be encrypted with the owner's public key, so that only the owner will be able to retrieve the actual data from the transaction. If the data is shared among multiple owners, a group key is used, as described in Section 3.4. Using this approach, an attacker will not be able to steal private information by eavesdropping on others' communications. In the search and rescue scenario, the attacker will not be able to steal data from the packets that he/she obtains a copy of, since the data transaction will be encrypted with the public key of the destinations. Hence, the attacker will not be able to decrypt it.

- Man in the middle: similar to the previous point, the attacker will not be able to modify the content of the transaction, since it will be encrypted with the destination's public key. If the transaction contains public data, then the attacker will be able to change the content. However, the attack will be detected by the receiver using the signature check. Since the attacker cannot generate the correct signature for the modified content (since the attacker doesn't have the private key of the generator), the signature of the transaction becomes invalid, and the receiver drops it. In the search and rescue scenario, the receiver of a packet that was maliciously modified by an attacker can detect the attack by checking the signature of the transaction inside the packet.

- 51% attack: in PROACT, the miners are the TGCSs. Hence, the attacker needs to compromise and control 51% of all the TGCSs to be able to insert malicious transactions into the BC. This is due to the fact that the consensus method proposed in [21] and adopted by PROACT depends on obtaining acknowledgments from 51% or more of the TGCSs to validate each transaction in the block. For example, suppose that we have 50 TGCSs in the network. When one of the TGCSs broadcasts its new block to the TGCS network, then the new block is added to the BC only after 26 TGCSs (51%) or more validate and acknowledge each transaction in the block. "Section 3.3. The proposed PoAT consensus" in [21] describes the detailed consensus mechanism in PoAT that was adopted in PROACT. Hence, the attacker needs to compromise 51% of the TGCSs to be able to add malicious transactions to the BC. This is a very hard task since TGCSs are distributed among various organizations/businesses (as described in Section 3.1) and are selected by the CAs as the most secured and trusted GCSs in each organization/company.

## 5. PERFORMANCE EVALUATION

We tested the performance of the proposed framework using the network simulation tool ns-3 (version 3.32). We created a custom ns-3 agent for the UAV node which includes the various operations that a drone should execute, as described in Section 3. We also implemented a custom model for the required blockchain system based on the basic bitcoin implementation (https://github.com/bitcoin/bitcoin). However, we modified the bitcoin code to add our own transaction and block types and parameters, encryption and signature operations, and storage methods (to allow storing part of the BC on drones). We also wrote our own implementation of the PROACT consensus protocol and integrated it into ns-3. In order to simulate the GCS, we adopted the QGroundControl (QGC) tool (https://github.com/mavlink/qgroundcontrol). We modified the QGC code to make the GCS receive commands from the CA, implement the proposed BC model, and execute the PROACT protocol. We simulated the connection between the UAVs and GCSs using the IEEE 802.16 standard (ns-3 WimaxNetDevice), and the connection between UAV nodes using the IEEE 802.11n standard (ns-3 WifiNetDevice). On the other hand, the connections between the CAs and between a CA and its GCSs were simulated using XG-PON links (updated ITU-T standard for Passive Optical Networks) [23].

The map that we used in the simulation scenarios is shown in Figure 4. This map represents a rural area that contains multiple large forests in the north of Lebanon. The size of the map is 1177 Km$^2$. We downloaded the ".osm" file of the map from *openstreetmap.org*. Next, we exported the map to the simulation of urban mobility (SUMO) software and we converted the SUMO simulation to the mobility and trace files of ns-3 using the SUMO *traceExporter* tool. We simulated a fire emergency scenario in which multiple CAs are deployed over this large area. Each CA is connected to several GCSs, and each GCS manages a number of UAVNs that will be used to identify the exact locations of fires, and the existence of injured or fire-trapped people. In addition, we simulated two sensors on each drone: a visual sensor for image capturing and a thermal sensor to detect the locations of fire and people. In the simulation scenarios, we make each drone send a data packet to the GCS every 2 seconds with a default size equal to 10KB. However, we present a section in which we studied the effect of large data transactions on the system performance. The number of CAs was set to 5 and the number of GCSs per CA was set to 10. The default number of UAVNs per GCS was set to 10 and the default number of UAVs per UAVN was set to 30. However, we include a section in which we varied the number of UAVNs per GCS between 5 and 20 and the number of UAVs per UAVN between 10 and 50. Figure 4 illustrates the locations of the CAs and GCSs on the utilized map. Among the 10 GCSs of each CA, 4 GCSs were selected as TGCSs (a total of 20 TGCAs). Note that the locations of the fixed nodes (CAs and GCSs) were defined in the ns-3 simulation script based on the map locations shown in Figure 4. These locations were selected based on the distribution of forests on the map. In the areas that contain no or few forests, less number of GCSs were deployed; while in areas that contain a large number of forests, a larger concentration of GCSs was deployed to cover all the forests in the area. The remaining simulation parameters are shown in Table 1.

**Table 1.** Simulation parameters

| Parameter | Value |
|---|---|
| Simulation area | $39.76 \times 29.6$ Km$^2$ |
| Scenario simulation time | 1 hour |
| Number of UAVs | 2500 – 50,000 (default 15,000) |
| Maximum drone speed | 20 – 30 m/s |
| Transmission range (UAV) | 300 m |
| Transmission range (GCS) | 1000 m |
| Number of CAs | 5 |
| Number of GCSs per CA | 10 |
| Data Transaction generation interval | 2s |
| Data Transaction size | 1 - 100 KB |
| Malicious attack rate | 10s |
| UAV total energy | 3600 kJ |
| Routing protocol | SURFER [20] |

We simulated five types of transactions which are:

1) $T_1$: configuration data and commands sent by the GCS to the UAV, which was simulated by making each GCS send a control packet to each drone in its UVANs every 1 sec. The size of the control packet was set to 100 bytes.

2) $T_2$: command sent to multiple UVAs. Each GCS selects a group of UAVs between 5 and 10 in one of its UAVNs and sends to them a multi-UAV command every 10s. The size of this transaction is also 100 bytes.

3) $T_3$: data sent by the drone to the GCS. A data packet of size 10KB was sent by each drone to its GCS every 2s.

4) $T_4$: security-related data. Whenever a drone S detects that another drone D is requesting access to a transaction that it doesn't have access to, or that it is fabricating false information, then S sends a $T_4$ transaction to the GCS containing the details of the malicious behavior of D. The size of this transaction was set to 1KB.

5) $T_5$: CA commands to the GCS. Each CA sends a control packet to each of its GCSs every 100ms. The size of the control packet was set to 100 bytes.

$T_3$ was defined as a $Tran_{T_1}$ transaction, $T_1$; $T_4$; and $T_5$ were defined as $Tran_{T_2}$ transactions, while $T_2$ was defined as a $Tran_{T_3}$ transaction. On the other hand, $T_3$ and $T_5$ were defined as $Trans_{S_1}$ transactions, while $T_1$; $T_2$; and $T_4$ were defined as $Trans_{S_2}$ transactions. For $Trans_{S_1}$ transactions, we used the 256-bit ECC standard algorithm for the encryption/decryption and the SPONGENT-224 hash function [24] for signing the transaction. On the other hand, we divided $Trans_{S_2}$ transactions into two categories based on the drone mission duration. If the drone that is sending/receiving a $Trans_{S_2}$ transaction is participating in a mission whose duration is less than 10 minutes, the transaction will be classified as a $Trans_{S_2}-C_1$ transaction. Else, the transaction will be considered as a $Trans_{S_2}-C_2$ one. We used the SPONGENT-88 hash function for signing both categories of $Trans_{S_2}$ transactions. However, $Trans_{S_2}-C_1$ transactions were encrypted/decrypted using a 64-bit ECC curve (using the *shortECC* technique that was proposed in [25]), while $Trans_{S_2}-C_2$ transactions were encrypted/decrypted using a 128-bit ECC curve. This division was chosen as follows: we simulated drone missions having a random duration between 5 and 60 minutes (time between starting the drone and ordering it to return to base). We divided done missions into two categories as we explained. For both categories, we need to use an ECC curve that ensures that an attacker will never be able to deduce the private key within the mission duration. For the first period (10 minutes), we chose 64-bit ECC which is enough since it is currently impossible to break a 64-bit ECC in 10 minutes. For the second period, which is between 10 and 60 minutes, we used a 128-bit ECC curve since it is currently impossible to break it in 1 hour. Finally, for $Tran_{S1}$ transactions, we chose the standard 256-bit ECC which is highly secured against current brute-force attacks.

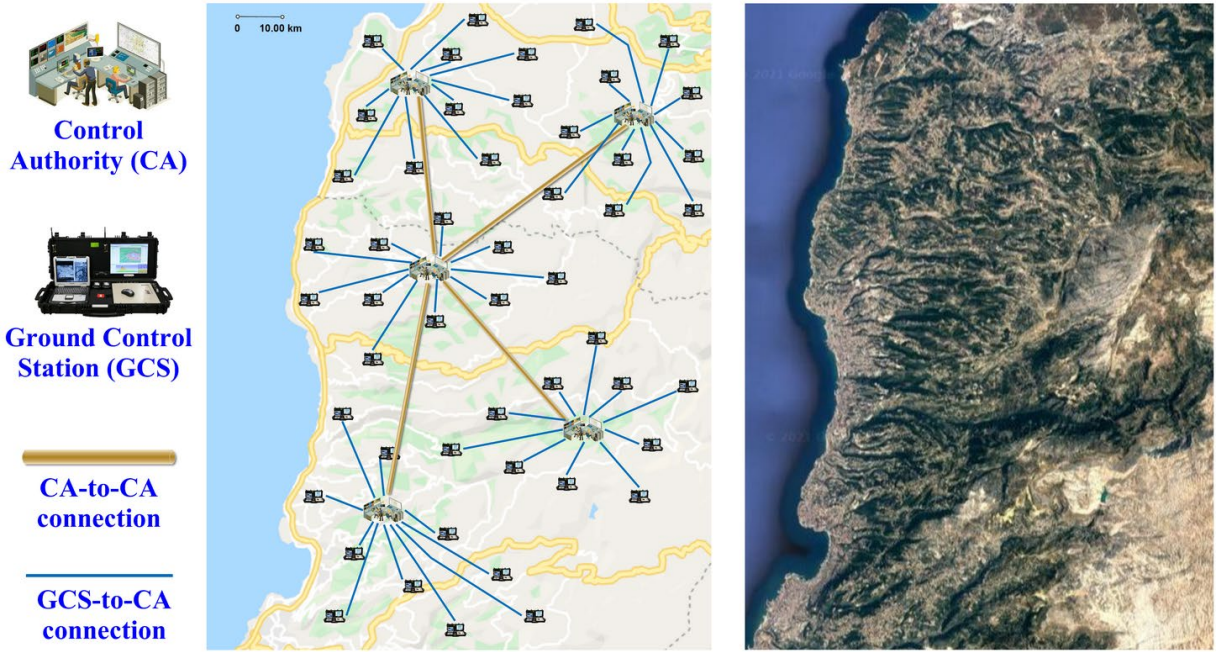

**Figure 4.** Map used in the simulations

We compare PROACT to two blockchain systems that were recently proposed for IoD, which are the Distributed Blockchain-based Framework for UAVs (DBFU) in [2] and the Drone-based Delegated Proof of Stake (DDPOS) [18]. The two systems were described in Section 2. Briefly, DBFU uses the Keccak function and limits the size of a transaction to 1KB, while DDPOS utilizes a zone-based model in which zone controllers handle the BC operations. The three protocols (DBFU, DDPOS, and PROACT) were compared by calculating the following four parameters for each protocol:

1) Attack Detection Rate (ADR): In each simulation scenario, we make a percentage ($M$) of the drones behave maliciously. Each malicious drone performs a malicious attack every 10 seconds by either sending to another drone a request to obtain a transaction in its BC that the malicious drone doesn't have the access right to, or by sending a data packet that contains false data to the GCS, such as the coordinates of an injured person or the location of a fire while there is no injured person or no fire (the data can be detected as false by the GCS if other drones that are near to the location of the malicious drone send the correct data). In each simulation scenario, we measure the ADR, which is the percentage of malicious transactions that are detected by the GCSs and not added to the blockchain. The ADR reflects the ability of the legitimate drones and the GCSs to detect the malicious requests and the false data that are sent by the malicious drones. The default value of $M$ was set to 0.2 (i.e., 20% of all drones are malicious). However, we present a section in which $M$ was varied between 10% and 70%.

2) Transaction-Blockchain Delay (TBD): which is the time between a transaction is generated by a source (which could be a drone, GCS, or CA) until the transaction is added to a new block in the BC. We measure the TBD of each transaction during the simulation scenario and at the end, we calculate the average for all transactions. The TBD parameter is very important as it denotes the time that the receiver or user should wait after the data is generated before being able to process the data from the BC. As denoted before, IoD delay-sensitive applications request a small TBD in order to perform correctly and efficiently.

3) Drone Energy Consumption (DEC, in kilojoules (kJ)): In order to measure the energy consumption per drone, we used the ns3 Energy Framework that was published in [26]. We modified the parameters of the "Energy Source" and "Device Energy Model" classes (we replaced some parameters and added other parameters) to include the parameters that affect the energy consumption of a drone. Our modifications were based on the UAV model that was proposed in [27]. The two main parameters that we varied and that play the major role in the energy consumption due to the blockchain system are $E_{processors}$ and $E_{communications}$ (the reader can refer to [27] for the details of the various parameters that affect the energy consumption). The $E_{processors}$ parameter was calculated based on the drone's total processing time at a certain instant, while $E_{communications}$ depends on the number of communications made by the drone and the duration of each communication. We assumed standard values for the drone's weight, speed, flight height, direction change, and other similar parameters, based on the calculations made by Abeywickrama et al. in [28]. We initiated the Energy Source of each drone to have a total energy density equal to 1000Wh (watt-hour) at the beginning of each simulation scenario, which is equivalent to 3600 kJ (kilojoules). This parameter denotes the maximum amount of energy that a drone can consume during the simulation scenario.

4) Blockchain Transaction Overhead (BTO): in order to show the effect of the lightweight BC mechanism that we proposed, we calculate the average overhead of each transaction within the block body. This parameter was calculated for all transactions that are saved in $Block_{T1}$ blocks (since our aim is to reduce the storage space on the drone). Suppose that the original size of the transaction is $S_{TO}$, and the size of the same transaction in the BC is $S_{TB}$ (which includes the parameters described in Section 3.6; the signature; and the transaction data after encryption), then the BTO of the transaction is calculated as $(S_{TB} - S_{TO})/S_{TO}$. The BTO was calculated for each $Block_{T1}$ transaction in each drone and the average was taken for all transactions.

The four parameters (ADR, TBD, DEC, and BTO) were calculated while varying the total number of UAVs in the network between 2500 and 50000 (Section 5.1), varying the data transaction size between 1 and 100KB (Section 5.2), and varying the percentage of malicious UAVs between 10% and 70% (Section 5.3).

## 5.1 Varying the Number of UAVs

As stated in the previous section, the number of CAs was set to 5 and the number of GCSs per CA was set to 10. Figure 4 shows the locations of the CAs and GCSs on the map. In the next two sections, we fix the number of UAVNs per GCS ($N_{UAVN/GCS}$) to 10 and the number of UAVs in each UAVN ($N_{UAV/UAVN}$) to 30, which results in a total number of 15,000 UAVs in the network. However, in this section, we study the effect of varying the total number of UAVs in the network ($N_{UAV}$) on the performance of the three systems. Hence, we simulated 11 scenarios, in which we vary $N_{UAVN/GCS}$ and $N_{UAV/UAVN}$. The values of $N_{UAVN/GCS}$, $N_{UAV/UAVN}$, and $N_{UAV}$ in each scenario are shown in Table II. For each scenario, we record the values of the four parameters (ADR, TBD, DEC, and BTO) for each of the three compared systems. The results are shown in Figure 5.

The ability of the GCSs and legitimate UAVs to detect the attacks of malicious UAVs increases as $N_{UAV}$ increases. This observation is common for the three systems, as Figure 5a illustrates. The reason is that as $N_{UAV}$ increases, the number of legitimate UAVs increases more than the number of malicious UAVs (since the default value of $M$ was set to 0.2). This increases the probability of a malicious UAV sending an attack request to a legitimate UAV (the attack will be undetected if the malicious UAV sends a malicious request to another malicious UAV). In addition, with a higher $N_{UAV}$, there is a higher probability that one of the legitimate UAVs is near to the location of the malicious UAV. In such case, the legitimate UAV sends correct data to the GCS while the malicious UAV sends wrong data, which enables the GCS to detect the attack. The ADR of PROACT is slightly higher than those of DBFU and DDPOS, which reflects the efficiency of the proposed mechanism in detecting malicious behavior. Since each UAV establishes a virtual connection with the GCS and sends the packets to the GCS encrypted with the latter's public key, then the GCS will be able to receive the security-related transactions and correct data transactions from legitimate UAVs and detect the attacks of malicious UAVs.

**Table 2.** Parameters of the simulation scenarios for Section 5.1

| Scenario | $N_{UAVN/GCS}$ | $N_{UAV/UAVN}$ | $N_{UAV}$ |
|---|---|---|---|
| 1 | 5 | 10 | 2500 |
| 2 | 10 | 10 | 5000 |
| 3 | 15 | 10 | 7500 |
| 4 | 7 | 30 | 10,500 |
| 5 | 10 | 30 | 15,000 |
| 6 | 13 | 30 | 19,500 |
| 7 | 10 | 50 | 25,000 |
| 8 | 12 | 50 | 30,000 |
| 9 | 15 | 50 | 37,500 |
| 10 | 17 | 50 | 42,500 |
| 11 | 20 | 50 | 50,000 |

Figure 5b shows the great advantage of the proposed parallel multi-miner consensus mechanism in decreasing the average transaction-to-blockchain delay (TBD). Enabling miners to produce their blocks in parallel and then add them to the BC according to the order defined by the BO decreases the average delay required to save a transaction in the BC from 670ms (DDPOS) and 860ms (DBFU) to 110ms (PROACT). This decrease in TBD is very important for delay-sensitive applications. In addition, the TBD of PROACT is less affected by the increase of $N_{UAV}$. While the TBD of DDPOS increases from 0.38 to 0.97 seconds and that of DBFU increases from 0.52 to 1.16 seconds, the TBD of PROACT increases from 0.07 to 0.14s.

Similar to the TBD, the DEC of PROACT is much less than those of DBFU and DDPOS (Figure 5c). As stated in the previous section, two main factors affect the DEC which are the total processing time and the number and duration of communications (the other factors that affect the DEC are the same for the three compared systems). In PROACT, the total processing time decreases with the lightweight cryptography mechanisms that are applied by the drone to encrypt and sign transactions. In addition, the process of saving part of the BC on the drone enables the drone to obtain the data transactions that are important to its operations without the need to connect to the GCS or to another drone, which highly reduces the communications of each drone. On the other hand, DBFU makes the drone save only the transactions whose size is less than 1KB, while other transactions are saved on the GCS BC, which leads to more communications since the drone will request such transactions from the GCS whenever needed. As for DDPOS, the BC is saved on the zone controllers only, which forces the drone to request each transaction that it needs from the zone controller's BC, and hence it has the highest number of communications between the drone and its GCS.

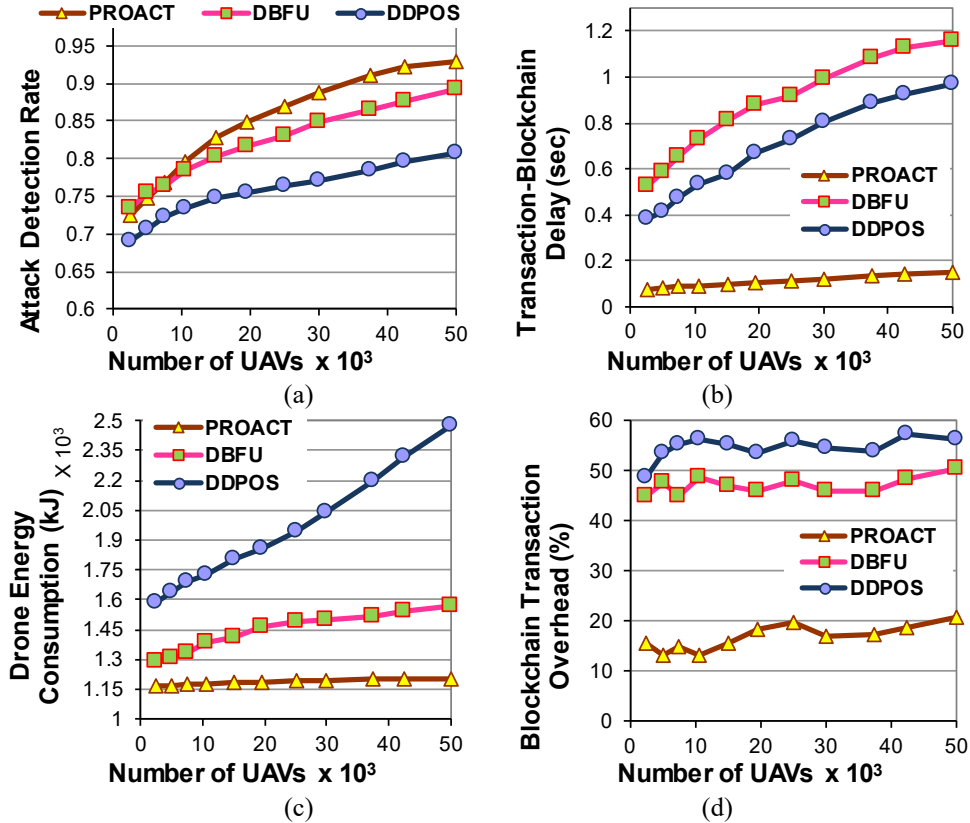

**Figure 5.** ADR, TBD, DEC, and BTO of the three systems while varying the number of UAVs between 2500 and 50,000

The last parameter that we study is the BTO, which is the percentage of added size to the transaction within the BC as compared to the original transaction data. Figure 5d shows that the BTO of PROACT varies between 13.2 and 20.6% as $N_{UAV}$ increases from 2500 to 50,000. On the other hand, the BTO of DBFU varies between 45 and 50.6% while that of DDPOS varies between 48.6 and 57.2%. Hence, we can deduce that PROACT decreases the overhead size of a transaction by an average of 34%, as compared to DBFU and DDPOS. This decrease is due to applying lightweight cryptography methods for signing and encrypting each $Trans_{S2}$ transaction. This decrease per transaction plays an important role in reducing the overall size of the block, which enables the drone to store more blocks within its limited storage space. This factor is very important for the successful implementation of a BC mechanism within the IoD. We also notice from Figure 5d that DBFU has a smaller size overhead than DDPOS, since DBFU uses the Keccak function for signing the BC transactions, while DDPOS utilizes the traditional EEC method.

## 5.2 Varying the Size of Data Transactions

Each UAV sends to the GCS a data transaction that contains simulated sensors' readings every 2s. The default size of the data transaction ($S_{DT}$) was set to 10KB. In this section, we study the effect of varying $S_{DT}$ between 1 and 100KB on the performance of the three systems. Figure 6a shows that the ADRs of the three systems decrease as $S_{DT}$ increases. In general, as $S_{DT}$ increases, more packets are required to send the transaction (since the maximum size of a packet is fixed). The increase in the number of packets sent per transaction increases the overall traffic in the network, which consequently leads to higher congestion, packet loss, and drop rate. This affects all types of messages sent by the drones to the GCS, including the security-related transactions ($T_4$), which causes the GCSs to detect a lower number of attacks. This issue is common for the three systems. The ADRs of PROACT and DBFU are similar when $S_{DT}$ is high.

The effect of the network congestion that occurs due to the higher number of packets that are required to send a transaction can be also observed on the TBD and DEC of the three systems. With a higher number of packets per transaction and higher network congestion, there is a higher probability that one or more packets will be dropped or delayed, which requires the drone to resend these packets. This will increase the delay required by the transaction to reach the GCS, and consequently increases the TBD. However, DBFU and DDPOS are more affected by the increase in $S_{DT}$ since they utilize traditional consensus mechanisms that require broadcasting new BC blocks to all BC nodes that participate in the consensus process. In PROACT, we limit the consensus to a smaller set of trusted GCSs (TGCSs). This method decreases the number of broadcast messages required per block and consequently leads to less network congestion and less time to reach consensus. Hence, the overall TBD of PROACT is less affected by the large transaction size as compared to that of DBFU and DDPOS.

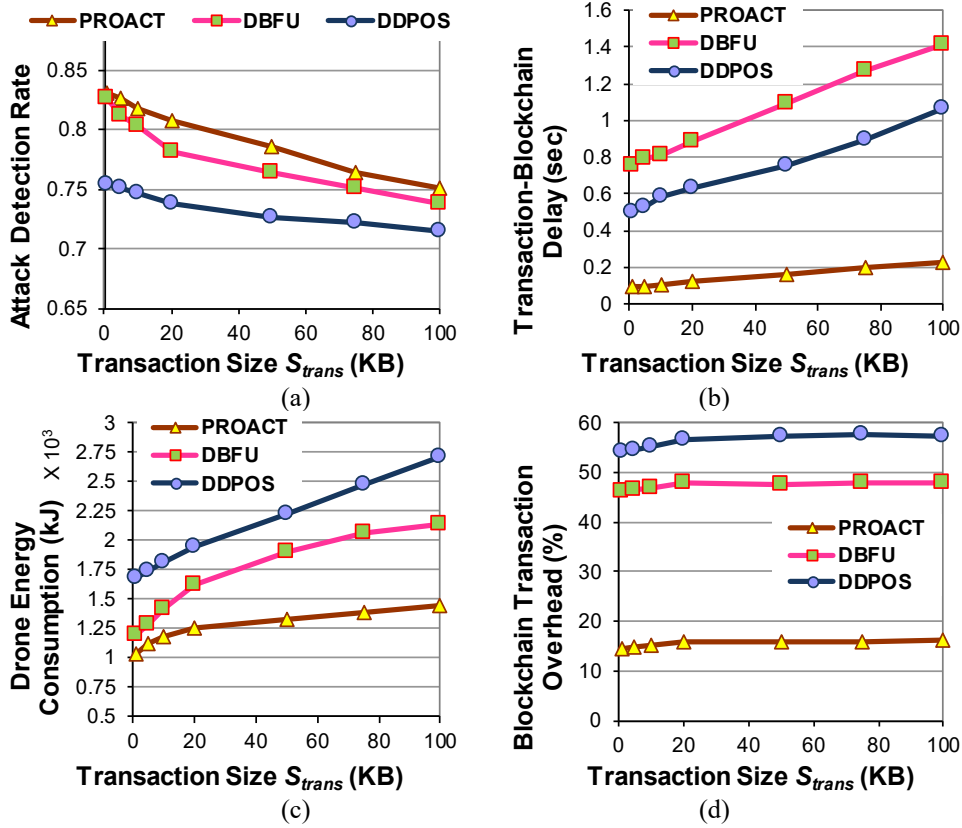

**Figure 6.** ADR, TBD, DEC, and BTO of the three systems while varying the size of data transactions between 1KB and 100KB

The DEC of the three systems also increases with $S_{DT}$ (Figure 6c). When $S_{DT}$ is small, DBFU and DDPOS have similar DECs, while the DEC of DDPOS is high (due to the fact that DDPOS requires a larger number of communications between the drone and the GCS, as explained in the previous section). As $S_{DT}$ increases, the DEC of PROACT increases but at a slower rate than those of the other two systems. This is also due to the reason stated in the previous section: a drone in PROACT obtains transactions from its BC, while a drone in DBFU and DDPOS sends requests to obtain some (DBFU) or all (DDPOS) required transactions from the GCS. As the size of the transaction increases, a larger number of packets are required per transaction. Hence, a larger number of communications should be made between the drone and the GCS. This increases the DEC of DBFU and DDPOS. The increase in the DEC of PROACT is due to the increased processing required to encrypt/sign the transaction that contains the generated data that the drone sends to the GCS (since a larger transaction will require more time to encrypt/sign and hence more energy consumption). However, the DECs of DBFU and DDPOS increase too much since the drone will consume more energy in both processing and communications.

As for the transaction overhead, Figure 6d shows that the BTOs of the three systems are not much affected by the increase in the $S_{DT}$. This is logical since the BTO is measured for all transactions within $Block_{T1}$ blocks, while the data transactions are saved in $Block_{T2}$ blocks. Hence, the change in $S_{DT}$ affects only some of the $Block_{T2}$ transactions and doesn't have any effect on $Block_{T1}$ transactions.

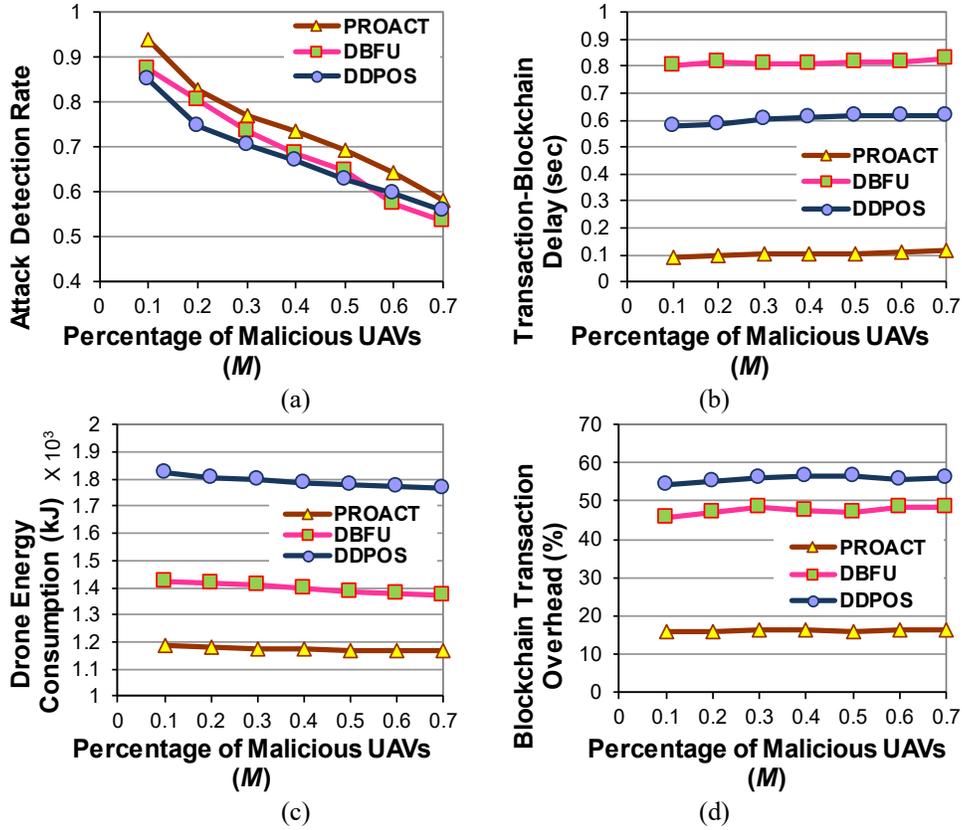

**Figure 7.** ADR, TBD, DEC, and BTO of the three systems while varying the percentage of malicious UAVs between 10 and 70%

## 5.3 Varying the Percentage of Malicious UAVs

The percentage of malicious UAVs ($M$) was set to a default value equal to 0.2 in the previous sections. In this section, we vary $M$ between 0.1 and 0.7 and observe the effect on the three systems. Figure 7a shows that the ADR of the three systems decreases significantly as $M$ increases. This is expected since the increase in $M$ means an increase in the number of malicious drones and a decrease in the number of legitimate ones. When the number of malicious drones increases, the number of cases in which malicious drones cooperate to perform attacks increases (such as a malicious drone sending an illegitimate request to another malicious drone). In addition, with the decrease in the number of legitimate drones, the number of cases in which a legitimate drone is near to a malicious drone and sends correct data that negates that of the malicious drone to the GCS decreases. For both reasons, the legitimate drones and the GCS become less able to detect malicious attacks as the number of legitimate drones decreases. Hence, the overall ADR decreases. Figure 7a shows that the three systems have similar ADRs when $M$ is very high.

While Figures 7b and 7d illustrate that the TBD and BTO are not much affected when $M$ is varied, Figure 7c shows that the DECs of the three systems slightly decrease with the increase of $M$. This is mainly due to the decrease in the communications between the drones and the GCS. As stated in the previous paragraph, when the number of legitimate drones decreases, the number of cases in which a legitimate drone detects a malicious request attack decreases. Hence, the number of packets that contain security-related transactions (i.e., $T_4$) decreases. The decrease in the number of these packets sent per drone causes a slight decrease in the DEC of the three systems.

## 5.4 Comparison with an Urban Area Scenario

The results that were described in the previous three sections were obtained by simulating a "search and rescue" scenario in a rural area. In this section, we compare with a similar scenario that was simulated in an urban area. For this purpose, we downloaded from *openstreetmap.org* a map that includes the greater city of Beirut and its suburbs, and we simulated a similar "search and rescue" scenario in which UAVNs are deployed, after an earthquake hits the city, to discover injured or trapped people based on analyzing the readings of thermographic sensors that were simulated in ns3. Figure 8 illustrates the locations of the CAs and GCSs on the map. Note that several differences exist between urban-based and rural-based IoD scenarios. For example, the density of UAVs (per unit area) in an urban scenario is usually much higher than that in a rural scenario. This was reflected in this section by varying the number of UAVs between 500 and 15,000, which corresponds to a UAV density between 15.1 and 453.2 drones/km$^2$, as compared to a density between 2.1 and 42.5 drones/km$^2$ in Section 5.1. Note that the area of the map in Figure 8 is equal to 33.1 km$^2$. In addition to the drones' density, the attack rate was increased in this section to 1 attack every 2 seconds instead of 10 seconds in the previous sections, since attacks in an urban area usually occur at a higher frequency. Finally, the size of the data transaction was also increased and set between [1, 200] KB, as compared to [1, 100] KB in Section 5.1. This is also a characteristic of urban scenarios: more applications exist and more data are exchanged.

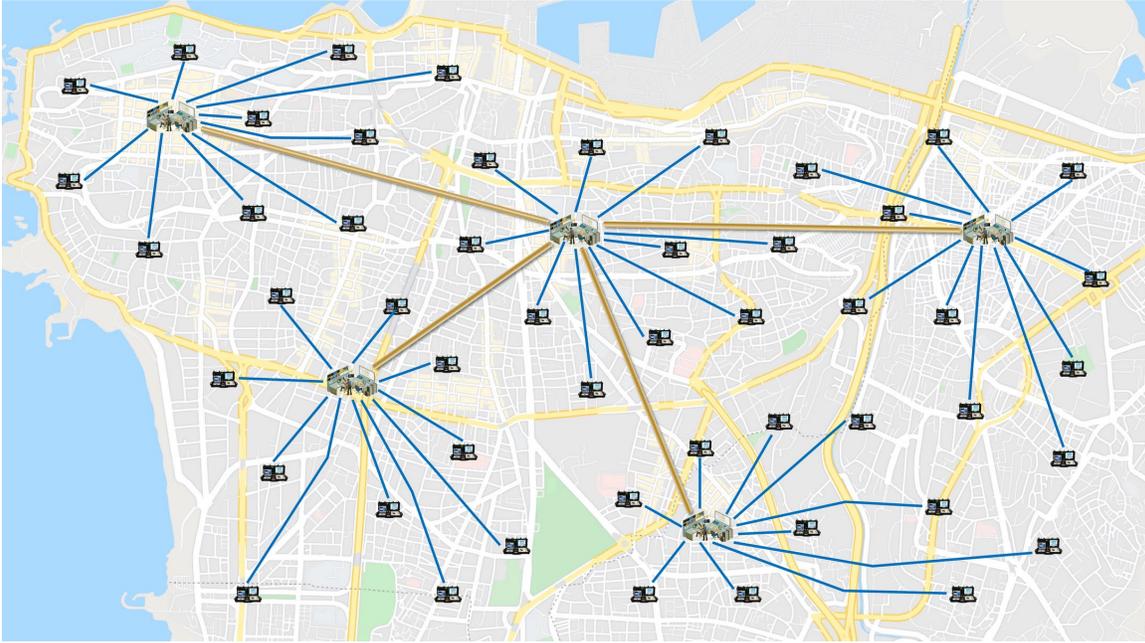

**Figure 8.** Map used in the simulation of the urban scenarios

Figure 9a illustrates the ADR of the three IoD systems in the urban scenario. We notice that the ADRs decrease, on average, as compared to the rural scenario. For example, the ADR of PROACT decreases from an average of 84% to 72.1%, while the ADR of DBFU decreases from 81.5 to 55.8%, and that of DDPOS decreases from 75.3 to 53.6%. The ADRs of the three systems maintain the same general behavior, which is that they increase as the total number of UAVs increases. However, the ADR of each scenario decreases as compared to the ADR of the rural scenario that contains the same number of UAVs. This is due to the increase in the drone's density, attack rate, and transaction size. These conditions make the IoD of the urban scenario more congested with nodes and packets, which affects the attack detection efficiency. In many cases, the GCS will depend on the correct data that it receives from the legitimate drones in order to detect the false data that is sent by a malicious drone. If the packet from the legitimate drone to the GCS is dropped or delayed too much due to network congestion, the attack will not be detected. We also notice that the ADR of PROACT is less affected by network congestion than the ADRs of DBFU and DDPOS.

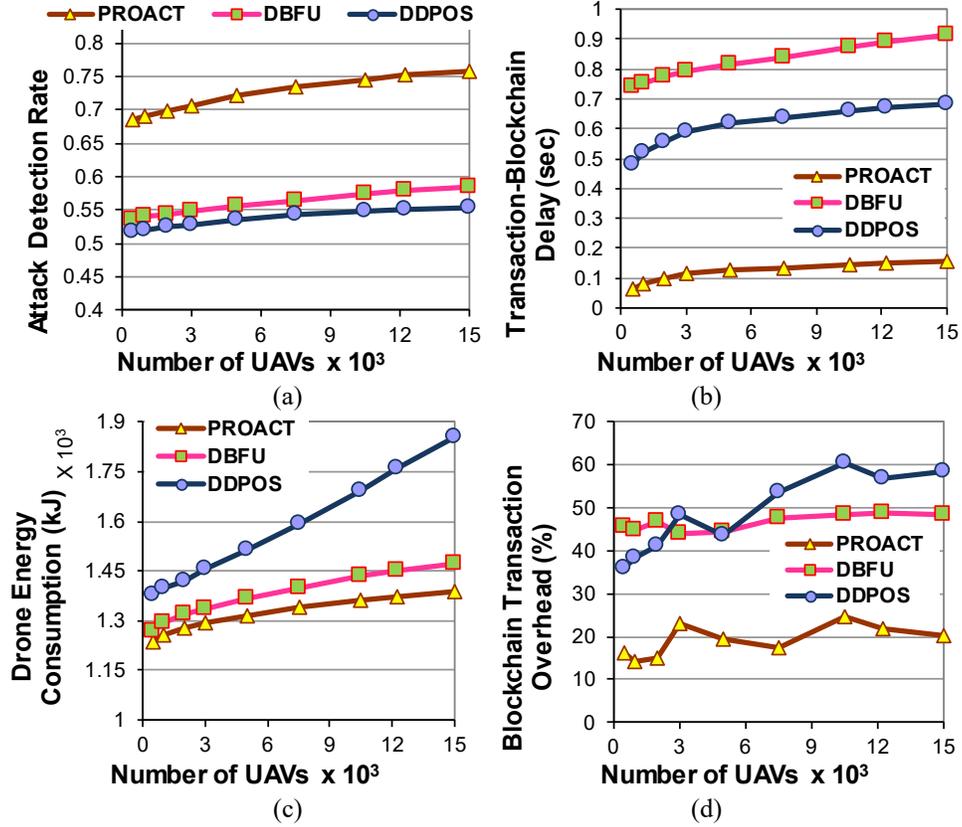

**Figure 9.** ADR, TBD, DEC, and BTO of the three systems in the urban scenario while varying the number of UAVs between 500 and 15,000

With respect to the transaction delay, we notice from Figure 9b that the TBD of PROACT increases very slightly as compared to the same TBD in the rural scenario. For example, the TBD of PROACT when the number of UAVs = 15,000 is equal to 148 ms in the rural scenario and 156 ms in the urban scenario. As for DBFU and DDPOS, their TBDs increase more than the TDB of PROACT. More specifically, the TBDs of DBFU and DDPOS increase in the urban scenario within a range of [98, 271] ms as compared to the rural scenario. On the other hand, the DECs of the three systems increase by comparable amounts, as we can observe in Figure 9c. From the figure, we can see that the DEC of PROACT increases from 1181 kJ in the rural scenario to 1386 kJ in the urban scenario when the number of UAVs is equal to 15,000. Similarly, the DEC of DBFU increases from 1415 to 1473 kJ, and that of DDPOS increases from 1807 to 1853 kJ. This indicates that the network congestion in the urban scenario has a considerable effect on the drone's energy consumption regardless of the implemented system. Finally, Figure 9d shows that the average BTO of the three systems remains approximately the same in the urban and rural scenarios. Although the BTO slightly changes between different scenarios, the average for all the scenarios of the same system is approximately the same. This is logical since the BTO depends on the cryptographic mechanisms that are used to sign and encrypt the transactions and hash the blocks, and is not affected by the number of nodes or the network congestion.

## 6. CONCLUSION AND FUTURE WORK

Safeguarding the data exchanged in the UAVN is vital for the success of IoD applications. While previous works attempted several approaches to implement the blockchain for this purpose, they fail to present a comprehensive solution that enables the UAVs to store and manage the BC in an efficient manner. In this paper, we proposed a BC model that takes into consideration the limited computational, storage, and energy capabilities of the UAV. Our system categorizes data transactions in the IoD based on their usage location and implements a mechanism that enables the UAV to save only the BC blocks that contain that data it needs. In order to reduce the computation and energy requirements of the BC operations, we proposed a novel method for implementing lightweight cryptography based on the required security duration of the transaction. Since many IoD applications are delay-sensitive and operate in real-time, it is crucial to save the data transactions in the BC as fast as possible after they are generated. For this reason, we presented PROACT, a consensus protocol that enables multiple miners to produce their blocks in parallel and add them to the BC based on the order specified by the block orderer. We compared the proposed framework with two recent BC systems and proved that our model excels in terms of block generation time, UAV energy consumption, and UAV storage requirements.

For future work, we will enhance the proposed system by integrating into it the capability to detect and defend against other IoD attacks, such as the Denial of Service (DoS) and navigation system-related attacks. We will also explore the possibility of integrating the PROACT protocol with the SDN model by separating the GCSs into data-plane GCSs that collect the data transactions and generate the BC blocks, and control-plane GCSs that validate the blocks and add them to the blockchain. This separation is intended to enhance the performance of the protocol and reduce the overhead on the TGCSs. Finally, we plan to augment the proposed BC framework with a machine learning model that will be used to monitor the IoD and adjust the system parameters (such as $TH_{CA}$, $T_{BIS}$, transaction security levels, etc.) in order to optimize the overall performance.

This will allow us to tune the values of the system parameters based on the performance criterion that is required by the IoD application (for example, reducing the TBD, DEC, BTO, etc.).

## REFERENCES

[1] M. Singh, G. S. Aujla, R. S. Bali, ODOB: One drone One block-based Lightweight Blockchain architecture for Internet of drones, in: 2020 IEEE Conference on Computer Communications Workshops, INFOCOM WKSHPS, IEEE, 2020, pp. 249-254.

[2] C. Ge, X. Ma, Z. Liu, A semi-autonomous distributed blockchain-based framework for UAVs system, *Journal of Systems Architecture* 107 (2020) 101728.

[3] E. Barka, C. A. Kerrache, H. Benkraouda, K. Shuaib, F. Ahmad, F. Kurugollu, Towards a trusted unmanned aerial system using blockchain for the protection of critical infrastructure, *Transactions on Emerging Telecommunications Technologies* 2019.

[4] H. Xu, W. Huang, Y. Zhou, D. Yang, M. Li, Z. Han, Edge computing resource allocation for unmanned aerial Vehicle Assisted mobile network with blockchain applications, *IEEE Transactions on Wireless Communications* 20 (5) (2021) 3107–3121.

[5] S. R. Pokhrel, Federated learning meets blockchain at 6G edge, in: Proceedings of the 2nd ACM MobiCom Workshop on Drone Assisted Wireless Communications for 5G and Beyond, DroneCom, ACM, 2020, pp. 49-54.

[6] S. Luo, H. Li, Z. Wen, B. Qian, G. Morgan, A. Longo, O. Rana, R. Ranjan, Blockchain-Based task offloading in drone-aided Mobile edge computing, *IEEE Network* 35 (1) (2021) 124–129.

[7] S. Aggarwal, M. Shojafar, N. Kumar, M. Conti, A new secure data Dissemination model in Internet of drones, in: Proceedings of the 53rd IEEE International Conference on Communications ICC, IEEE, 2019, pp. 1-6.

[8] M. Aloqaily, O. Bouachir, A. Boukerche, I. A. Ridhawi, Design guidelines for blockchain-assisted 5G-UAV networks, *IEEE Network* 35 (1) (2021) 64–71.

[9] N. Andola, Raghav, V. K. Yadav, S. Venkatesan, S. Verma, SpyChain: A Lightweight blockchain for authentication and Anonymous authorization in IOD, *Wireless Personal Communications* 119 (1) (2021) 343–362.

[10] B. Bera, A. K. Das, A. K. Sutrala, Private blockchain-based access control mechanism for unauthorized UAV detection and mitigation in Internet of Drones environment, *Computer Communications* 166 (2021) 91–109.

[11] T. Nguyen, R. Katila, T. N. Gia, A Novel Internet-of-Drones and Blockchain-based System Architecture for Search and Rescue, in: 18th IEEE International Conference on Mobile Ad-Hoc and Smart Systems, MASS, IEEE, 2021.

[12] Y. Tan, J. Liu, N. Kato, Blockchain-Based Key Management for Heterogeneous Flying Ad Hoc Network, *IEEE Transactions on Industrial Informatics* 17 (11) (2021) 7629–7638.

[13] M. Yahuza, M. Y. Idris, I. B. Ahmedy, A. W. Wahab, T. Nandy, N. M. Noor, A. Bala, Internet of drones security and Privacy Issues: Taxonomy and open challenges, *IEEE Access* 9 (2021) 57243–57270.

[14] M. S. Rahman, I. Khalil, M. Atiquzzaman, Blockchain-Powered Policy Enforcement for Ensuring Flight Compliance in Drone-Based Service Systems, *IEEE Network* 35 (1) (2021) 116–123.

[15] N. Pathak, A. Mukherjee, S. Misra, AerialBlocks: Blockchain-enabled UAV Virtualization for Industrial IoT, *IEEE Internet of Things Magazine* 4 (1) (2021) 72–77.

[16] S. Liao, J. Wu, J. Li, A. K. Bashir, W. Yang, Securing collaborative environment monitoring in smart cities using blockchain enabled software-defined internet of drones, *IEEE Internet of Things Magazine* 4 (1) (2021) 12–18.

[17] N. Hu, Z. Tian, Y. Sun, L. Yin, B. Zhao, X. Du, N. Guizani, Building agile and Resilient UAV networks based on SDN and blockchain, *IEEE Network* 35 (1) (2021) 57–63.

[18] A. Yazdinejad, R. M. Parizi, A. Dehghantanha, H. Karimipour, G. Srivastava, M. Aledhari, Enabling drones in the Internet of things with decentralized Blockchain-Based Security, *IEEE Internet of Things Journal* 8 (8) (2021) 6406–6415.

[19] M.Y. Arafat, S. Moh, A Q-learning-based topology-aware routing protocol for flying ad hoc networks, *IEEE Int. Things J.* 9(3) (2021) 1985–2000, https://doi.org/10.1109/JIOT.2021.3089759.

[20] K. Mershad, SURFER: A Secure SDN-Based Routing Protocol for Internet of Vehicles, *IEEE Internet of Things Journal* 8 (9) (2021) 7407–7422.

[21] K. Mershad, O. Cheikhrouhou, L. Ismail, Proof of accumulated trust: A new consensus protocol for the security of The IoV, *Vehicular Communications* 32 (2021) 100392.

[22] K. Mershad, H. Artail, Score: data scheduling at roadside units in vehicle ad hoc networks, in: 2012 19th International Conference on Telecommunications, ICT, IEEE, 2012, pp.1–6.

[23] X. Wu, K. Brown, C. Sreenan, P. Alvarez, M. Ruffini, N. Marchetti, D. Payne, L. Doyle, An XG-PON module for the ns-3 network simulator, in: Proceedings of the 6th International ICST Conference on Simulation Tools and Techniques, SimuTools, ACM, 2013, pp. 195-202.

[24] A. Bogdanov, M. Knežević, G. Leander, D. Toz, K. Varıcı, I. Verbauwhede, SPONGENT: A lightweight hash function, in: Proceedings of the 13th International Workshop on Cryptographic Hardware and Embedded Systems, CHES, Springer, 2011, pp. 312–325.

[25] A. Sojka-Piotrowska, P. Langendoerfer, Shortening the security parameters in lightweight WSN applications for IoT - lessons learned, in: 2017 IEEE International Conference on Pervasive Computing and Communications Workshops, PerCom Workshops, IEEE, 2017, pp. 636-641.

[26] H. Wu, S. Nabar, R. Poovendran, An Energy Framework for the Network Simulator 3 (ns-3), in: Proceedings of the 4th International ICST Conference on Simulation Tools and Techniques, SimuTools, ACM, 2011, pp. 222-230.

[27] U. C. Cabuk, M. Tosun, R. H. Jacobsen, O. Dagdeviren, A Holistic Energy Model for Drones, in: Proceedings of the 28th Signal Processing and Communications Applications Conference, SIU, IEEE, 2020, pp. 1-4.

[28] H. V. Abeywickrama, B. A. Jayawickrama, Y. He, E. Dutkiewicz, Comprehensive Energy Consumption Model for Unmanned Aerial Vehicles, Based on Empirical Studies of Battery Performance, *IEEE Access* 6 (2018) 58383–58394.